\documentclass[onecolumn]{aastex631}
\shorttitle{AASTeX v6.3.1 Sample article}
\shortauthors{Chainakun et al.}
\graphicspath{{./}{figures/}}

\usepackage[para,online,flushleft]{threeparttable}
\usepackage{color}

\begin{document}

\title{Mapping the X-ray corona evolution of IRAS~13224--3809 with the power spectral density}

\correspondingauthor{Poemwai Chainakun}
\email{pchainakun@g.sut.ac.th}

\author[0000-0002-9099-4613]{Poemwai Chainakun*}
\affiliation{School of Physics, Institute of Science, Suranaree University of Technology, Nakhon Ratchasima 30000, Thailand}
\affiliation{Center of Excellence in High Energy Physics and Astrophysics, Suranaree University of Technology, Nakhon Ratchasima 30000, Thailand}

\author[0000-0002-4516-6042]{Wasuthep Luangtip}
\affiliation{Department of Physics, Faculty of Science, Srinakharinwirot University, Bangkok 10110, Thailand}
\affiliation{National Astronomical Research Institute of Thailand, Chiang Mai 50180, Thailand}

\author[0000-0002-9639-4352]{Jiachen Jiang}
\affiliation{Institute of Astronomy, University of Cambridge, Madingley Road, Cambridge CB3 0HA, UK}

\author[0000-0003-3626-9151]{Andrew J. Young}
\affiliation{H.H. Wills Physics Laboratory, Tyndall Avenue, Bristol BS8 1TL, UK}

\begin{abstract}

We develop the power spectral density (PSD) model to explain the nature of the X-ray variability in IRAS~13224--3809, including the full effects of the X-ray reverberation due to the lamp-post source. We utilize 16 {\it XMM-Newton} observations individually as well as group them into three different luminosity bins: low, medium and high. The soft (0.3--1~keV) and hard (1.2--5~keV) PSD spectra are extracted and simultaneously fitted with the model. We find that the corona height changes from $h \sim 3 \ r_{\rm g}$ during the lowest luminosity state to $\sim 25 \ r_{\rm g}$ during the highest luminosity state. This provides further evidence that the source height from the reverberation data is significantly larger than what constrained by the spectral analysis. Furthermore, as the corona height increases, the energy spectrum tends to be softer while the observed fractional excess variance, $F_{\rm var}$, reduces. We find that the PSD normalization is strongly correlated with $F_{\rm var}$, and moderately correlated with the PSD bending index. Therefore, the normalization is dependent on accretion rate that controls the intrinsic shape of the PSD. While the intrinsic variability of the disk is manifested by the reverberation signals, the disk and corona may evolve independently. Our results suggest that, during the source height increases, the disk itself generates less overall variability power but more high-frequency variability resulting in the PSD spectrum that flattens out (i.e. the inner disk becomes more active). Using the luminosity-bin data, the hint of Lorentzian component is seen, with the peak appearing at lower frequencies with increasing luminosity. 

\end{abstract}


\section{Introduction} \label{sec:intro}

The active galactic nucleus (AGN) of the narrow-line Seyfert 1 galaxy IRAS~13224--3809 contains a maximally spinning supermassive black hole \citep{Fabian2013, Chiang2015, Jiang2018} with the mass of $\sim 2 \times 10^{6} M_{\odot}$ \citep{Alston2020}. It is one of the AGNs with a complex energy spectrum and complex X-ray variability over a broad range of timescales \citep[e.g.][]{Alston2019}. The energy-integrated spectrum during the long 2011 {\it XMM-Newton} observation could be explained by a patchy disk model that produced two reflection components from separate ionized elements, with an overabundance of iron \citep{Fabian2013}. The disk black-body emission and the narrow emission line at $\sim 6.4$~keV were required. Furthermore, the time delays between X-ray variability in the reflection and continuum dominated energy bands (referred to as reverberation lags) were measured by \cite{Kara2013}. They found that during low-flux stages, changes in the observed reverberation lags supported a compact coronal geometry.

\cite{Chiang2015} discovered that the reflection component of IRAS~13224--3809 is substantially less variable than the power-law emission, supporting the light-bending framework \citep[e.g.][]{Miniutti2004}. The spectral fitting revealed a constant emitting region that produced a soft thermal emission following the expected $L \propto T^{4}$ blackbody relation \citep{Chiang2015, Jiang2018}, which could have an accretion disk origin \citep{Ponti2010}. \cite{Chainakun2016} performed simultaneous modelling of the mean and lag-energy spectra taken into account the full dilution and ionization effects, and 
found that the soft excess was dominated by the narrow components which could be produced by the distant reflection from the cold torus, rather than dominated by the broad features from the inner disk reflection.
 
Furthermore, \cite{Parker2017} identified a series of variable peaks in the long-term X-ray variability spectra that could be interpreted as the strong absorption lines from an ultra-fast outflow (UFO). \cite{Alston2019} found that the power spectrum density (PSD) of IRAS~13224--3809 observed in 2002 ($\sim 64$~ks), 2011 ($\sim 500$~ks), and 2016 ($\sim 1.5$~Ms) contained non-stationary multiple-peaked components whose normalization increased while the low-frequency peak moved to higher frequencies when the source flux decreased. The intrinsic PSD shape is mostly determined by the mass accretion rate fluctuations that are propagated along the accretion disk \citep{Lyubarskii1997,Churazov2001,Arevalo2006,Ingram2011}. The accretion rate of IRAS~13224--3809 implied from the PSD spectra was comparable to that of black hole X-ray binaries in very-high/intermediate states \citep{Alston2019}. 

Recently, \cite{Jiang2022} utilized the high-density disk model to fit the time-averaged spectra of IRAS 13224–3809. They considered a broken power-law emissivity and a free reflection fraction parameter. Based on ionization and density parameters assuming the geometry is a lamppost, the source heights were calculated to be $\sim3$--6 gravitational radii ($r_{\rm g}=GM/c^{2}$; $M$ is the central balck hole mass, $G$ is the gravitational constant and $c$ is the speed of light). By considering the lag-frequency spectra in multiple observations during 2002--2016, the source height was found to increase with increasing luminosity, from $\sim 6 \ r_{\rm g}$ to $20 \ r_{\rm g}$ \citep{Alston2020}. Furthermore, \cite{Caballero2020} analyzed the combined spectral-timing data simultaneously in different flux periods. For the maximally spinning case, they also found a tendency of rising source height, from $\sim 3 \ r_{\rm g}$ to $10^{+10}_{-1} \ r_{\rm g}$, with luminosity.

Nevertheless, the PSD profiles can be imprinted with X-ray reverberation patterns \citep{Emmanoulopoulos2016, Papadakis2016, Chainakun2019a}, providing an independent way to probe the coronal geometry. Due to the gravitational light-bending effects, the reflection component is less variable compared to the power-law continuum. The reverberation signals then reduce the fractional excess variance ($F_{\rm var}$) so producing the dip in the PSD profiles that is more prominent in the more reflection-dominated band. The models for the X-ray reverberation signatures for the lag-spectra of AGN have been investigated extensively \citep[e.g.][]{Wilkins2013,Cackett2014,Emmanoulopoulos2014,Chainakun2016, Epitropakis2016}, especially in IRAS~13224--3809 \citep{Alston2020,Caballero2020}. We then choose to investigate the X-ray variability power using the PSD model that includes the full effects of the X-ray reverberation. We focus on the energy-dependence of the PSD shapes for 16 observations of IRAS~13224--3809 under the lamp-post assumption. We explore two energy bands: 0.3--1 keV and 1.2--5 keV as the representative of the reflection-dominated and continuum-dominated bands, respectively. This allows us to check for consistency the implied framework and lamp-post geometry using different timing profiles. 

In Section 2, we present the data reduction and how the PSD spectra are produced. The theoretical PSD models including  X-ray reverberation effects are described in Section 3. Section 4 explains the model grid created in this study as well as the fitting procedure. The best-fit results are presented in Section 5. Discussion and conclusion is given in Section 6 and 7, respectively.

\section{Observations and data reduction} 

\begin{table*}
\begin{center}
   \caption{\emph{XMM-Newton} observations of IRAS~13224--3809.} \label{tab:xmm_obs}
   \label{tab_obs}
   \begin{threeparttable}
    \begin{tabular}{lccccc}
    \hline
    Observation ID & Revolution number & Observational date & Exposure time$^{a}$ & Count rate$^{b}$ & Luminosity bin$^{c}$ \\
    & & & (ks) & (count s$^{-1}$) & \\
    \hline
0673580101	&	2126	&	2011-07-19	&	34.08	&	0.37	&	high	\\
0673580201	&	2127	&	2011-07-21	&	49.26	&	0.24	&	medium	\\
0673580301	&	2129	&	2011-07-25	&	52.03	&	0.09	&	low	\\
0673580401	&	2131	&	2011-07-29	&	85.11	&	0.28	&	medium	\\
0780560101	&	3037	&	2016-07-08	&	39.38	&	0.19	&	medium	\\
0780561301	&	3038	&	2016-07-10	&	112.35	&	0.26	&	medium	\\
0780561401	&	3039	&	2016-07-12	&	99.76	&	0.22	&	medium	\\
0780561501	&	3043	&	2016-07-20	&	93.22	&	0.13	&	low	\\
0780561601	&	3044	&	2016-07-22	&	97.85	&	0.38	&	high	\\
0780561701	&	3045	&	2016-07-24	&	100.13	&	0.16	&	low	\\
0792180101	&	3046	&	2016-07-26	&	110.64	&	0.13	&	low	\\
0792180201	&	3048	&	2016-07-30	&	110.55	&	0.19	&	medium	\\
0792180301	&	3049	&	2016-08-01	&	86.44	&	0.07	&	low	\\
0792180401	&	3050	&	2016-08-03	&	98.63	&	0.75	&	high	\\
0792180501	&	3052	&	2016-08-07	&	102.66	&	0.23	&	medium	\\
0792180601	&	3053	&	2016-08-09	&	101.50	&	0.68	&	high	\\
     \hline
     \end{tabular}
    \begin{tablenotes}
    \textit{Note.} $^{a}$Good exposure time after data cleaning. $^{b}$The background-subtracted source count rate in the 1--4 keV energy band and $^{c}$the luminosity bin in which the data belong to, regarding to its count rate (see text).  
    \end{tablenotes}
    \end{threeparttable}
    \end{center}
\end{table*}

The X-ray data of IRAS~13224--3809 used in this work were previously observed by {\it XMM-Newton} observatory \citep{Jansen2001} and were obtained from {\it XMM-Newton} Science Archive.\footnote{\url{http://nxsa.esac.esa.int}} Since one purpose of this study is to analyze each observational data individually, to get high signal to noise data, we selected only the observations which have the total exposure time of $\ga$100 ks. The {\it XMM-Newton} observational data used here are tabulated in Table~\ref{tab:xmm_obs}. To avoid the combination of data from pn and MOS detectors which were observed with different time resolutions, we chose to consider only the pn data which have higher time resolution and effective area.\footnote{\url{https://xmm-tools.cosmos.esa.int/external/xmm_user_support/documentation/uhb/epic.html}} We performed the data reduction using Science Analysis Software (SAS) version 19.1.0 with the latest version of the calibration files (CCF).\footnote{\url{https://www.cosmos.esa.int/web/xmm-newton/download-and-install-sas}} The pn observation data files were reprocessed using the SAS task {\sc epproc} with the default parameter values. Then, the observational periods which were affected by high background flaring activity were also removed by the SAS task {\sc espfilt} using the method {\it histogram} with the  parameter {\it allowsigma} of 2.5 (default value). The remaining exposure time after removing background flaring for each observation is shown in the fourth column of Table~\ref{tab:xmm_obs}.

For all observations, we consider the lightcurves of IRAS~13224--3809 extracted in two energy bands which are 0.3--1 keV (reflection dominated) and 1.2--5 keV (continuum dominated) referred to as the soft and hard energy bands, respectively.  The background-subtracted light curve in each energy band was extracted from the events flagged with PATTERN $\leqslant$ 4 and \#XMMEA\_EP using the SAS task {\sc evselect} and {\sc epiclccorr}; the source extraction region was defined as a circular area centred at the source position with the radius of 20 arcsec, while the background region was defined as a 60 arcsec radius circle, located on a source-free area that is still on the same CCD chip with that of the source region. Then, the produced light curves were converted into the power spectral density (PSD) utilising the {\sc ftools} task {\sc powspec};\footnote{\url{https://heasarc.gsfc.nasa.gov/lheasoft/ftools/fhelp/powspec.txt}} concisely, each light curve was divided into a number of segments with the length and the time-bin resolution of 20 ks and 179 s, respectively. Then all segments were converted into the PSDs, in which their output frequency ($f$) is corresponding to 0.05 -- 2.8 mHz, and averaged over; the Poisson noise was also subtracted during this step. Finally, the obtained PSD was rebinned logarithmically, in which the width of the next, higher frequency bin is larger by a factor of 1.06 ($f$ $\rightarrow$ 1.06$f$), to get the PSD for analysing further.

Moreover, we investigated the variability of PSD as a function of the luminosity when increasing their signal to noise.  To do this, we categorised the observational data into three groups -- low, medium and high luminosities -- based on the source's luminosity; here we used the background-subtracted, instrumental count rate, i.e. that of the pn detector, of the source in the 1--4 keV energy band as a proxy of the luminosity since this band should be dominated by the primary X-ray emission \citep{Caballero2020}. The low, medium and high luminosity observations were defined as the observations that have the count rate $<$ 0.19 counts s$^{-1}$, 0.19 counts s$^{-1}$ $\leqslant$ count rate $\leqslant$ 0.28 counts s$^{-1}$, and count rate $>$ 0.28 counts s$^{-1}$, respectively. The count rate and the group that each observation belongs to are shown in column 5 and 6 of Table~\ref{tab:xmm_obs}. The light curves in each luminosity bin were grouped together and then converted in the single PSD of each luminosity bin following the method explained above. The PSDs obtained from individual observations and grouped observations were then used in our analysis.

Note that the common 1--4 keV energy band was used only for grouping the observational data into 3 different luminosity states: low, medium, and high luminosity. When we calculated the PSD in the soft and hard bands, we selected to follow \cite{Alston2020} where the 0.3--1 keV and 1.2--5 keV bands were used to represent the soft (reflection dominated) and the hard (intrinsic emission dominated) bands, respectively.

\section{Variability power model}

The PSD of AGN can be generally explained by a broadband power-law with one or two bend frequencies where the profile changes its slope \citep{Papadakis2004,Gonzalez2012}. We use the PSD model in the form of a bending power-law \citep[e.g.][]{Emmanoulopoulos2016}:
\begin{equation}
P_{\rm 0}(f) = A f^{-1}\bigg{[}1+ \bigg{(}\frac{f}{f_{b}} \bigg{)}^{s-1}\bigg{]}^{-1} \; ,
    \label{psd1}
\end{equation} 
where the PSD has a low-frequency slope $-1$ that is bent gradually to the high-frequency slope $s$ above the bending frequency $f_{\rm b}$. $A$ is the normalization factor of the variability power.

The reverberation signals imprint the oscillatory structures on the intrinsic PSD profile \citep{Papadakis2016, Chainakun2019a}. We use the ray-tracing simulations to generate the response functions of the disk reflection under the lamp-post scenario, by tracing photon paths along the Kerr geodesics from the source to the disk and to the observer's sky \citep[e.g.][]{Karas1992, Fanton1997, Wilkins2013, Cackett2014, Chainakun2016, Emmanoulopoulos2014, Epitropakis2016, Caballero2018, Caballero2020}. The observer is stationary at $1000 \ r_{\rm g}$ from the black hole. The X-ray reprocessing by the disk is calculated via the {\sc reflionx} model \citep{George1991, Ross1999,Ross2005}, where the photon index and the iron abundance are fixed at $\Gamma=2.4$ and $A_{\rm Fe}=15$, respectively \citep{Chainakun2016}. The disk extends from the inner-most stable circular orbit (ISCO) to $400 \ r_{\rm g}$. The black hole spin is fixed at $a=0.998$ \citep{Fabian2013, Chiang2015, Jiang2018}, so the ISCO is at $\sim 1.23 \ r_{\rm g}$. The disk response function is produced by collecting the flux of the reflected photons as a function of time at the observer's sky.  

Let us assume that $\Psi(f, E_{j})$ is the response function for the X-ray reverberation scheme, the observed PSD including reverberation effects can be computed via \cite{Uttley2014, Papadakis2016, Chainakun2019a, Chainakun2021b} as
\begin{equation}
P_{\rm rev}(f) \propto |\Psi(f)|^{2} P_{\rm 0}(f) \; ,
    \label{psd}
\end{equation}
where $P_{\rm 0}(f)$ is the intrinsic PSD profile expressed in eq.~\ref{psd1}. Therefore, $P_{\rm 0}(f)$ acts as a driving signal that is filtered by the response function, causing the observable dip and oscillatory structure in the $P_{\rm rev}(f)$. These echo features are more prominent in a more reflection-dominated band. We normalize the area under the response function to 1 and employ the reflection fraction $R=$ (reflection flux)/(continuum + reflection) to reduce the effect of the energy band on the PSD profiles \citep[e.g.][]{Chainakun2019a}, which is otherwise responsible for an energy-dependent amount of dilution applied to the reverberation calculations. This also accounts for the uncertainty in the variations of, e.g., continuum photon index and iron abundance among different individual observations which are not constrained here. 

Additional narrow features in the broadband PSD profiles were previously suggested in IRAS~13224--3809 \citep{Alston2019}, to resemble the observed PSD shape of very-high/intermediate state black hole X-ray binaries \citep{Remillard2006}. We model these narrow features using the Lorentzian function given in the form of:
\begin{equation}
P_{\rm lor}(f) = \frac{N (\sigma_{\rm lor}/2 \pi)}{(f - f_{\rm lor})^{2} + (\sigma_{\rm lor}/2)^{2} }  \; ,
    \label{lor}
\end{equation}
where $N$ is a normalization factor, $\sigma_{\rm lor}$ is the FWHM of the Lorentzian line and $f_{\rm lor}$ is the centroid frequency of the line. The PSD data are fitted using the $P_{\rm rev}$ model, or the ($P_{\rm rev} + P_{\rm lor}$) model if the additional Lorentzian component is required by the data. 
 
\section{Model grid and fitting procedure}
\label{sec:model_grid&fitting}
The response functions corresponding to the source height $h$ between 2--40~$r_{\rm g}$ are simulated with equal spacing of the model grid of $2 \ r_{\rm g}$. We vary the bending power-law index $s$ in between 1--4 with the grid spacing of 0.2, $f_{\rm b}$ in between (1--10)$\times 10^{-4}$~Hz with the spacing of $10^{-4}$~Hz, and $R$ in between 0.1--0.8 with the spacing of 0.1. To avoid the model degeneracy, we fix the $M=2 \times 10^{6} M_{\odot}$ \citep{Alston2020} and the inclination $i=45^{\circ}$ \citep{Caballero2020}. 

For each observation, the source height $h$ is tied between the soft (0.3--1 keV) and hard (1.2--5 keV) band data sets, while $s$, $f_{\rm b}$, $R$, and $A$ are allowed to be free. Our parameters then consist of the source height, $h$, which is tied between two energy bands, the PSD bending index ($s_{\rm s}$ and $s_{\rm h}$), bending frequency ($f_{\rm b, s}$ and $f_{\rm b, h}$), reflection fraction ($R_{\rm s}$ and $R_{\rm h}$), and the PSD normalization ($A_{\rm s}$ and $A_{\rm h}$) in the soft and hard bands for all the parameters. Different combinations of these parameter values represent different grid cells of the model.

We perform the simultaneous fits of the soft and hard PSD data in {\sc isis} \citep{Houck2000} by stepping through the model grid cell. The $\chi^2$ statistics are calculated using the {\sc subplex} optimization method. The best-fit parameters are those that match to the grid cell with the lowest $\chi^2$ value. Finer local grids are created if necessary, and the fitting is repeated to acquire the new best-fit parameter values.

\section{Results}

\begin{figure*}
\centerline{
\includegraphics*[width=0.55\textwidth]{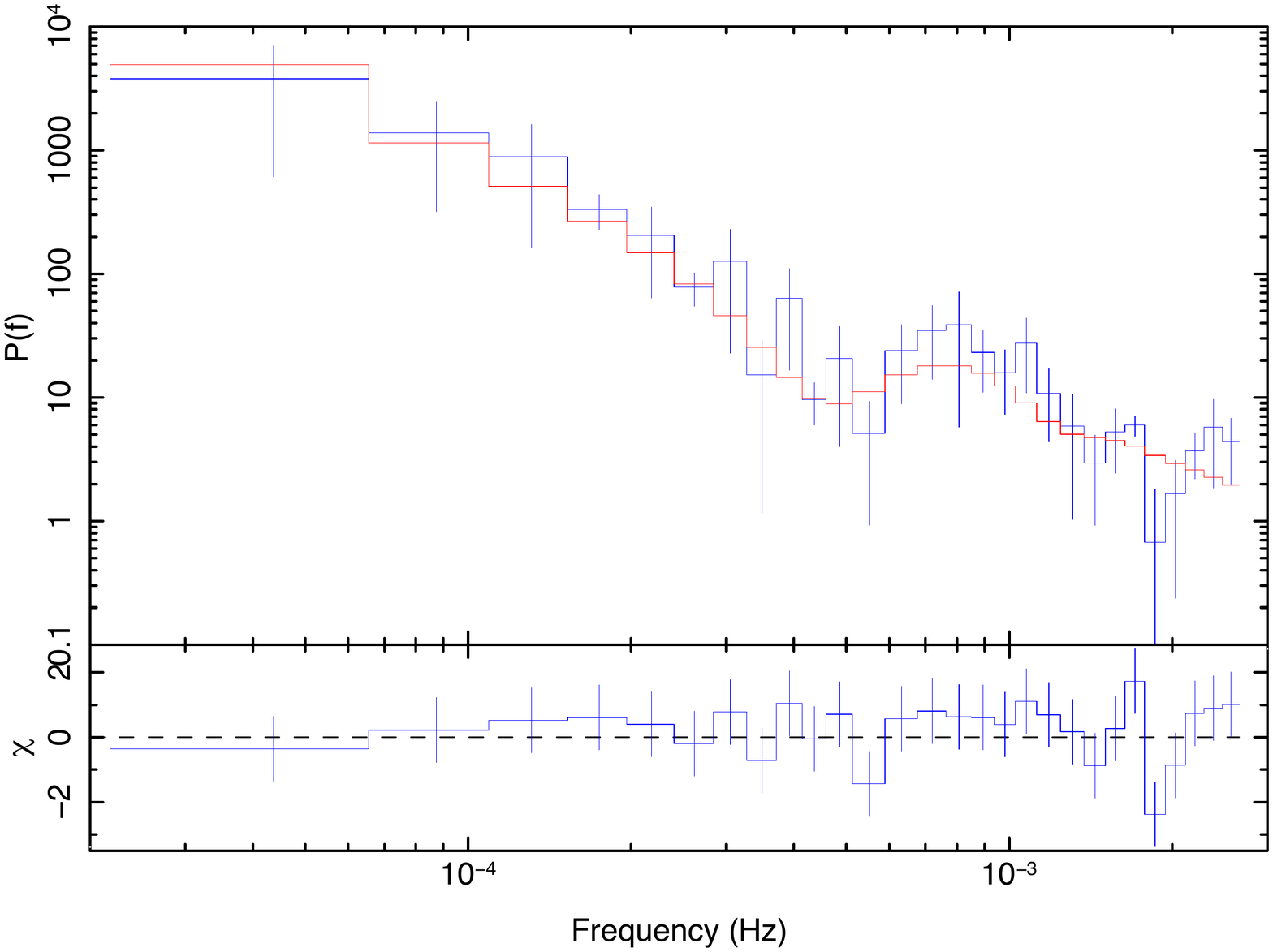}
 \put(-100,175){Rev. no. 2127}
  \put(-100,165){0.3--1 keV}
\hspace{-1.0cm}
\includegraphics*[width=0.55\textwidth]{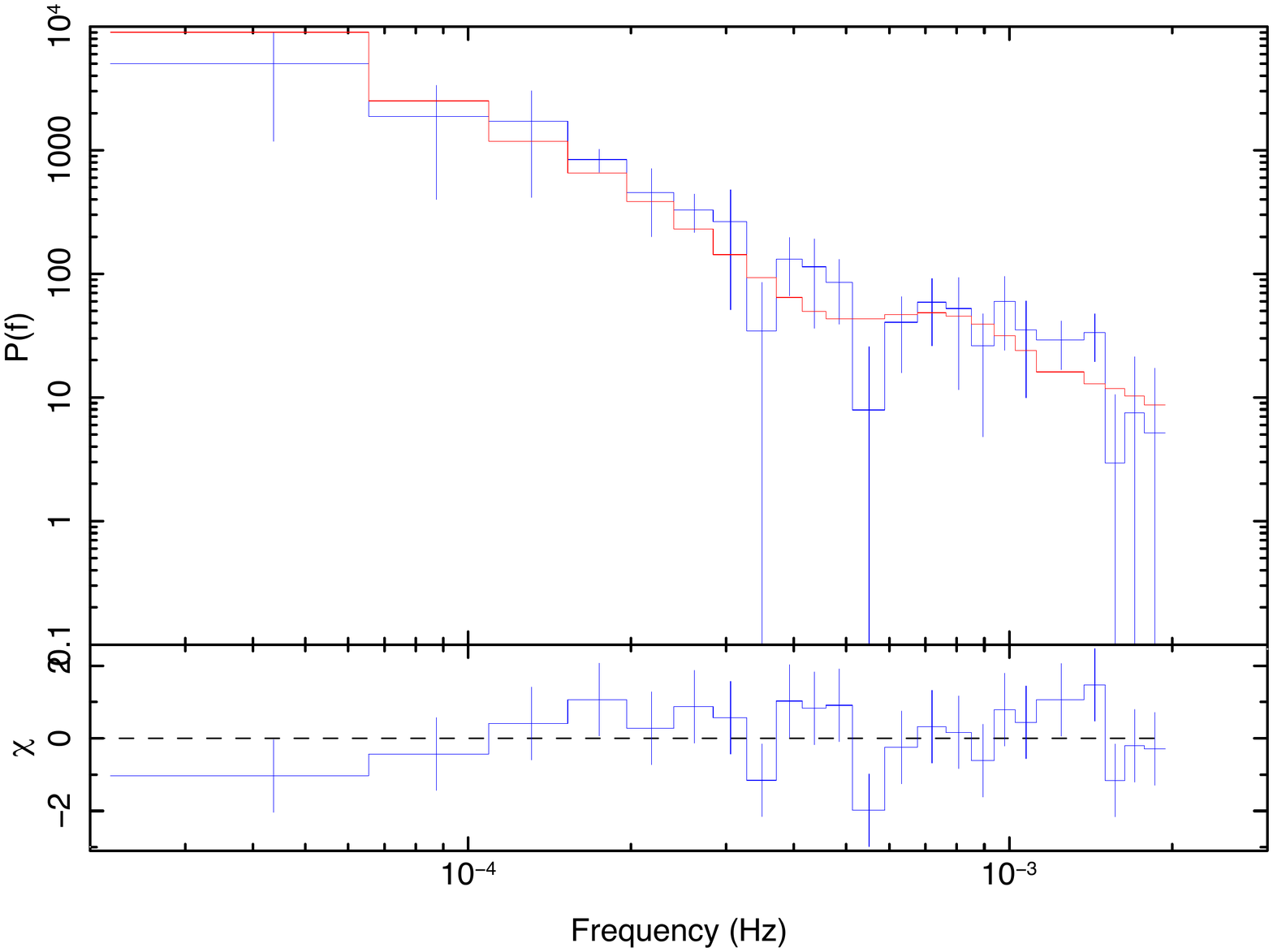}
 \put(-100,175){Rev. no. 2127}
  \put(-100,165){1.2--5 keV}
\vspace{-1.0cm}
}
\centerline{
\includegraphics[width=0.55\textwidth]{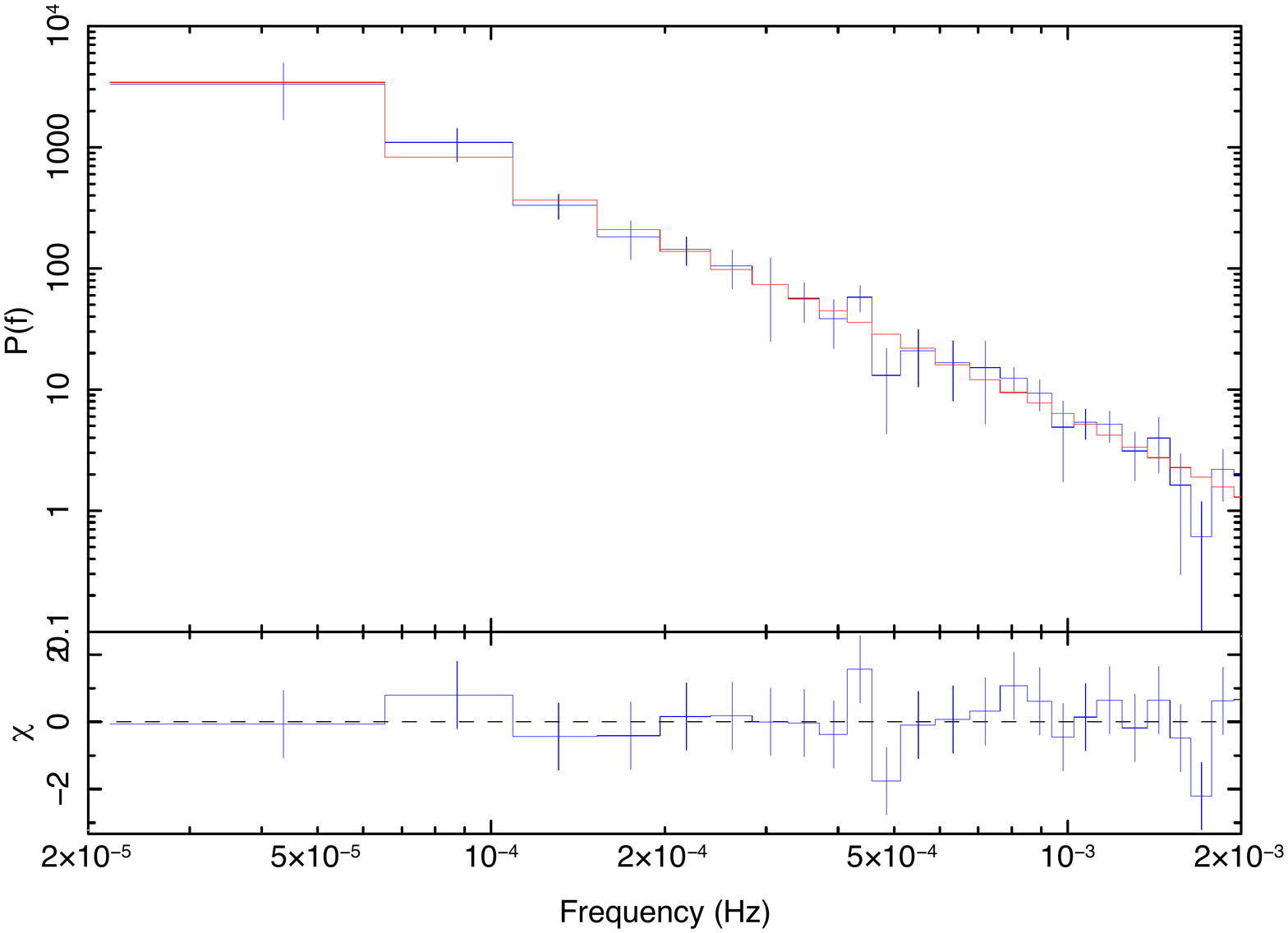}
 \put(-100,175){Rev. no. 2131}
  \put(-100,165){0.3--1 keV}
\hspace{-1.0cm}
\includegraphics[width=0.55\textwidth]{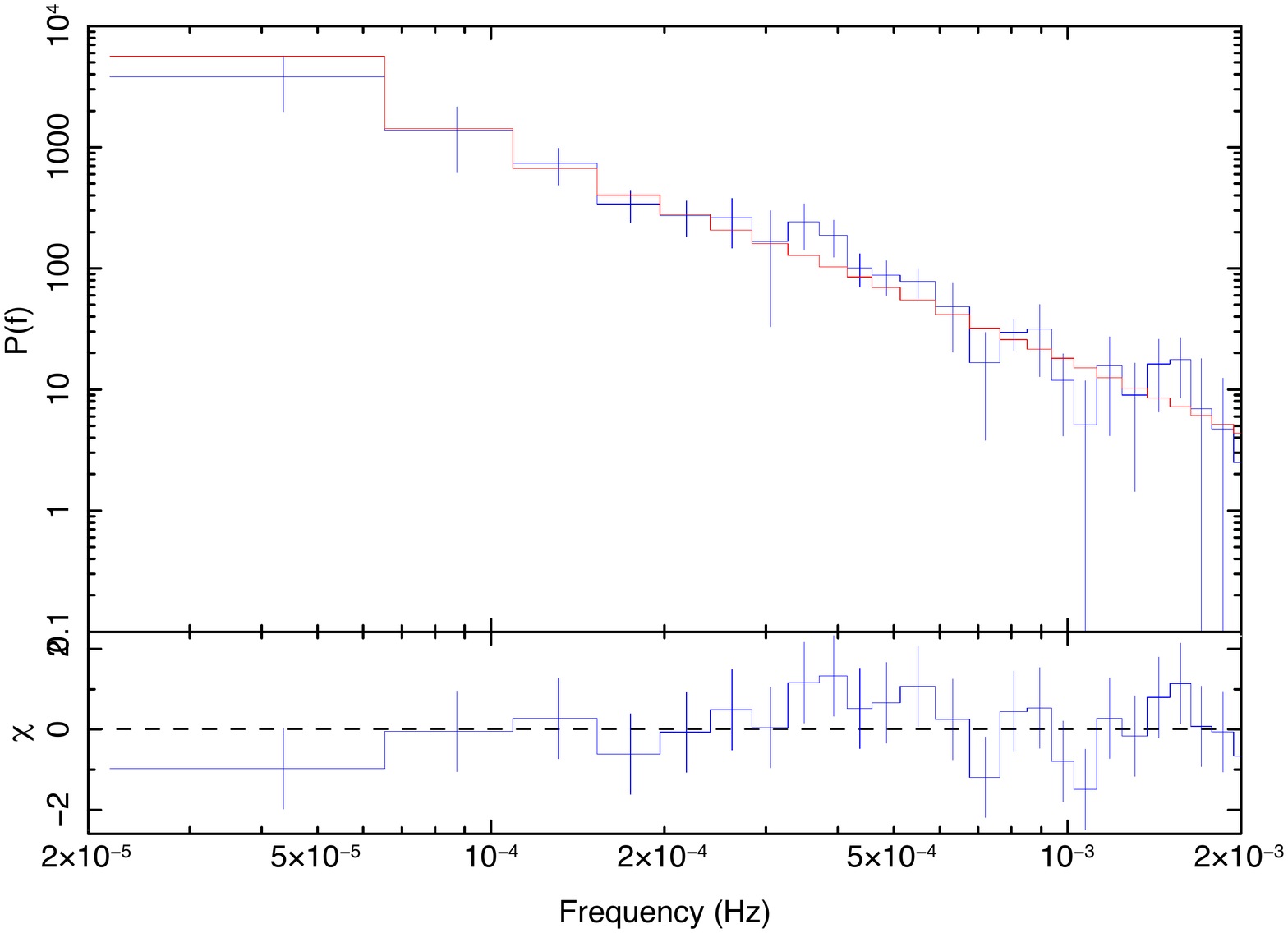}
 \put(-100,175){Rev. no. 2131}
  \put(-100,165){1.2--5 keV}
\vspace{-1.0cm}
}
\centerline{
\includegraphics[width=0.55\textwidth]{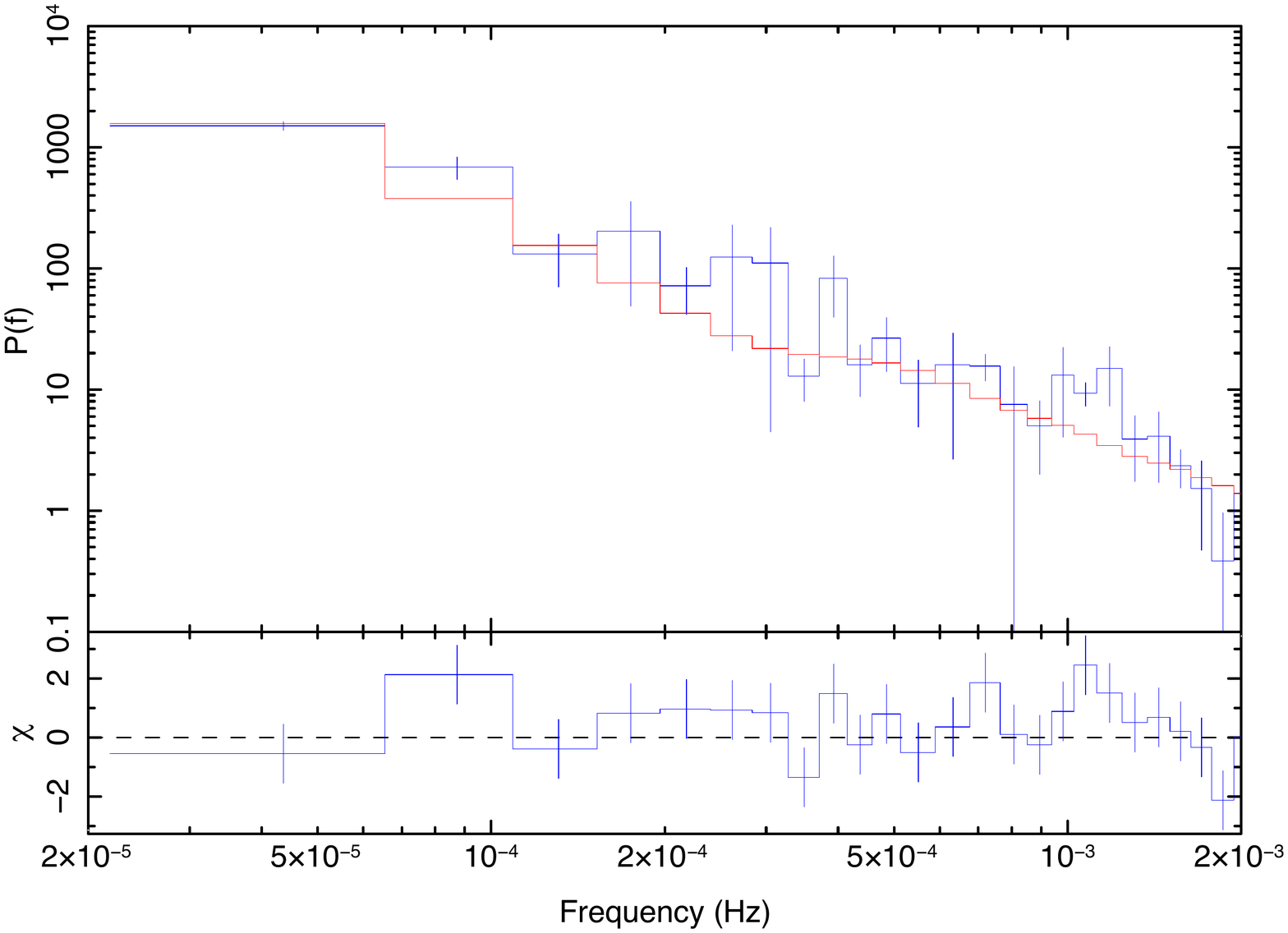}
 \put(-100,175){Rev. no. 3037}
  \put(-100,165){0.3--1 keV}
\hspace{-1.0cm}
\includegraphics[width=0.55\textwidth]{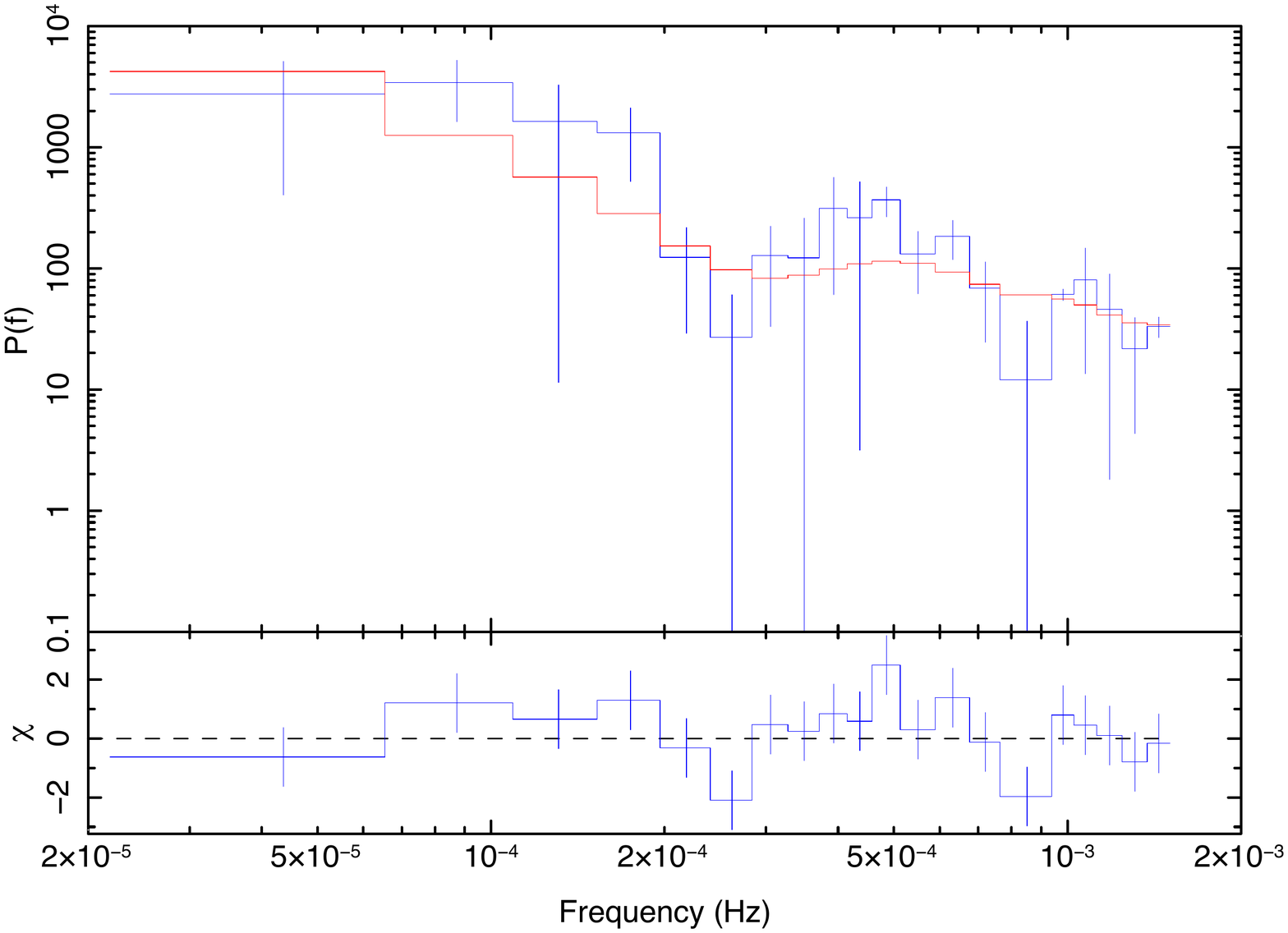}
 \put(-100,175){Rev. no. 3037}
  \put(-100,165){1.2--5 keV}
}
\caption{Data, model and residuals from simultaneously fitting the $P_{\rm rev}$ model (red) to the Poisson noise-subtracted PSD data (blue) in the 0.3--1 keV band (left panels) and 1.2--5 keV band (right panels) for different individual observations. The black hole mass is fixed at  $2 \times 10^{6} M_{\odot}$. }

\end{figure*}

\addtocounter{figure}{-1}
\begin{figure*}
\centerline{
\includegraphics*[width=0.55\textwidth]{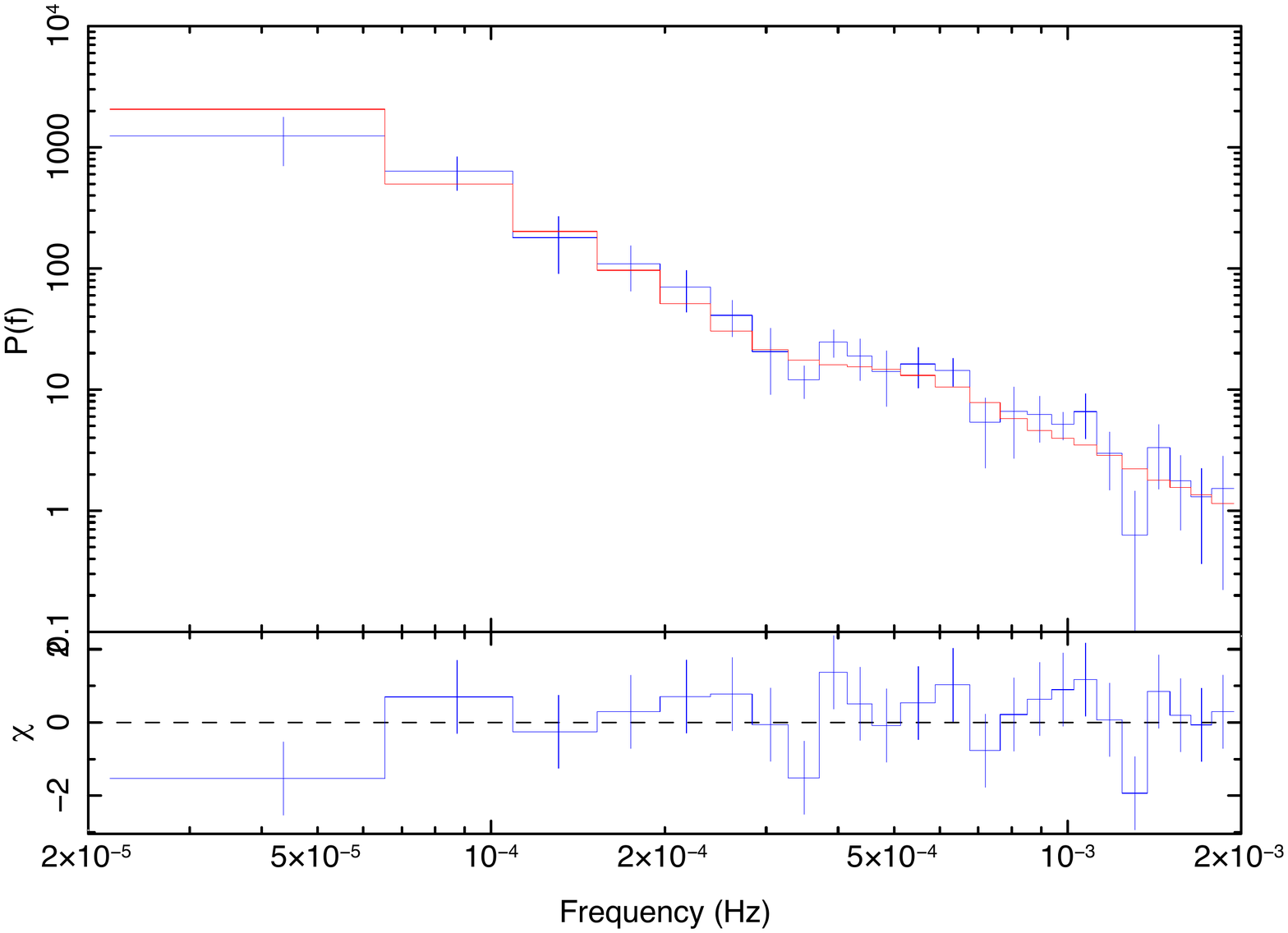}
 \put(-100,175){Rev. no. 3038}
  \put(-100,165){0.3--1 keV}
\hspace{-1.0cm}
\includegraphics*[width=0.55\textwidth]{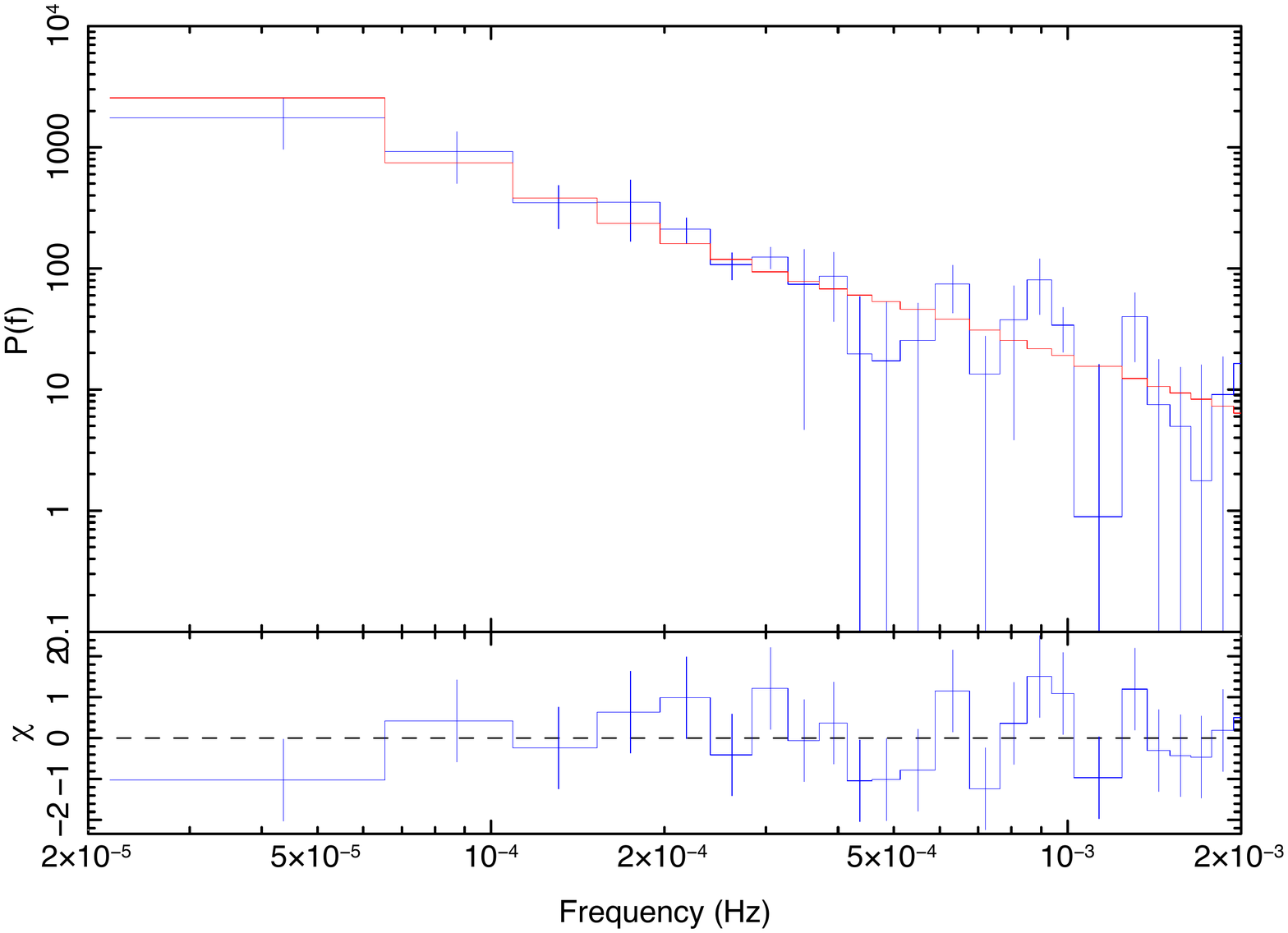}
 \put(-100,175){Rev. no. 3038}
  \put(-100,165){1.2--5 keV}
\vspace{-1.0cm}
}
\centerline{
\includegraphics[width=0.55\textwidth]{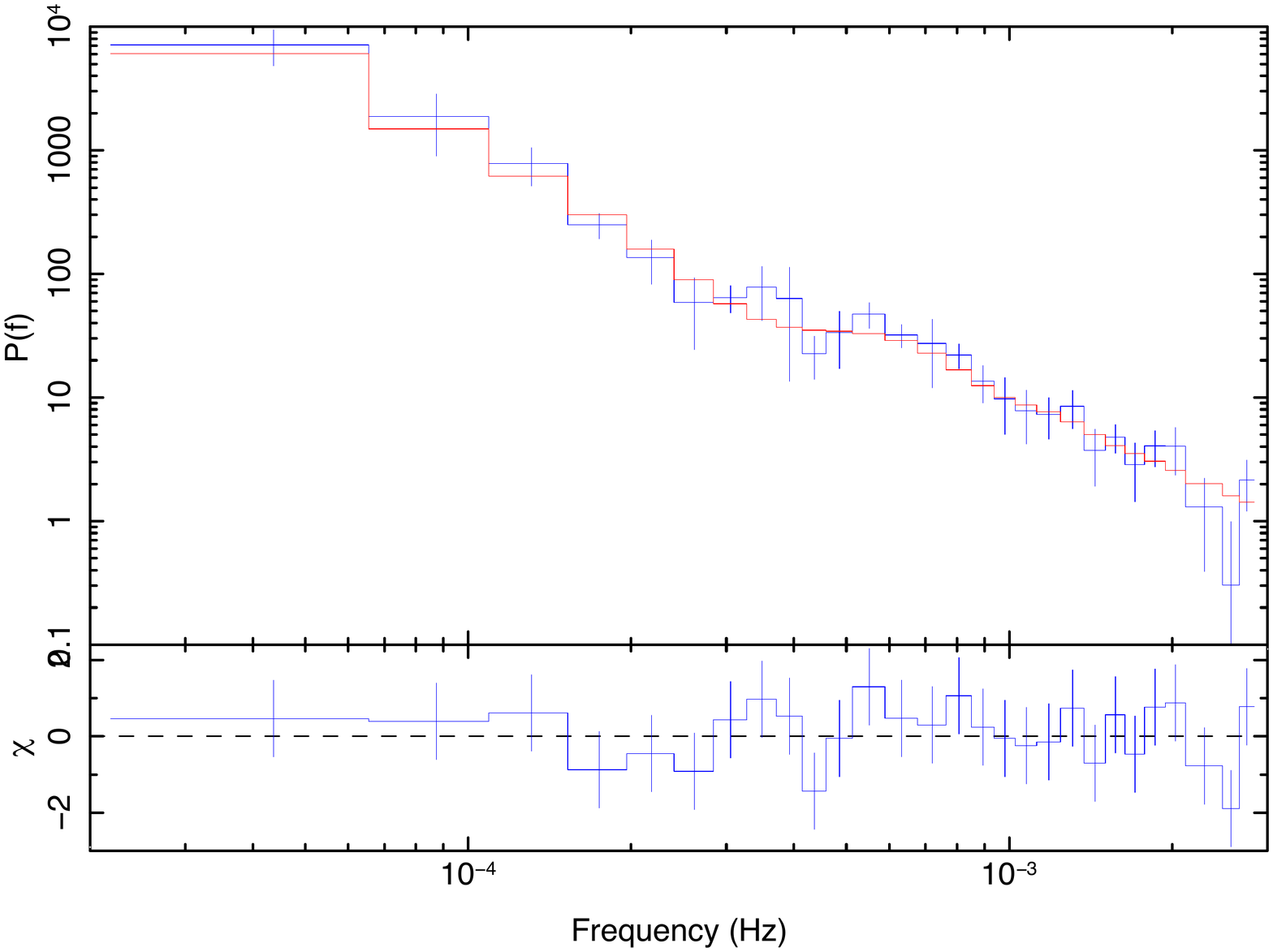}
 \put(-100,175){Rev. no. 3046}
  \put(-100,165){0.3--1 keV}
\hspace{-1.0cm}
\includegraphics[width=0.55\textwidth]{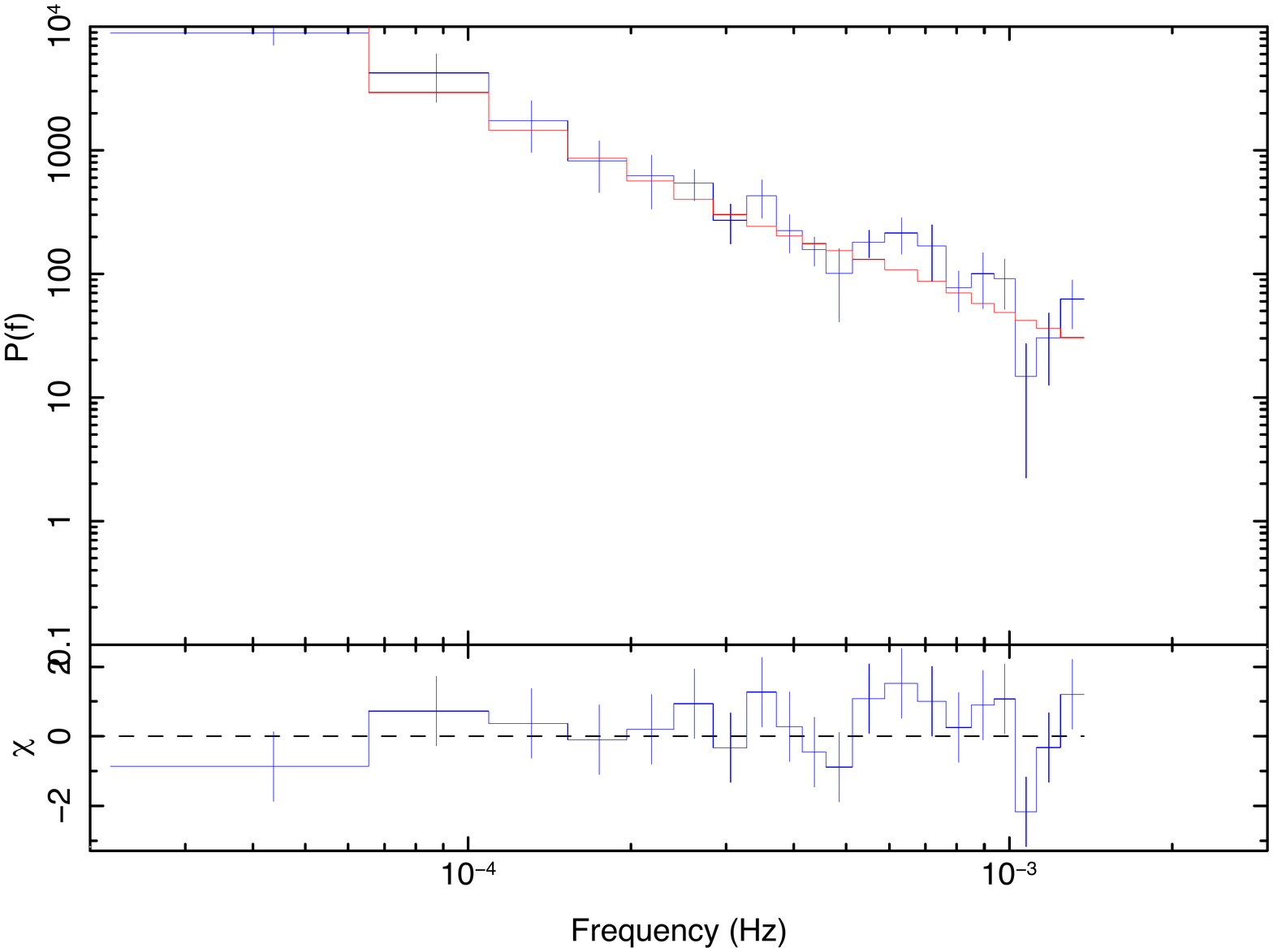}
 \put(-100,175){Rev. no. 3046}
  \put(-100,165){1.2--5 keV}
\vspace{-1.0cm}
}
\centerline{
\includegraphics[width=0.55\textwidth]{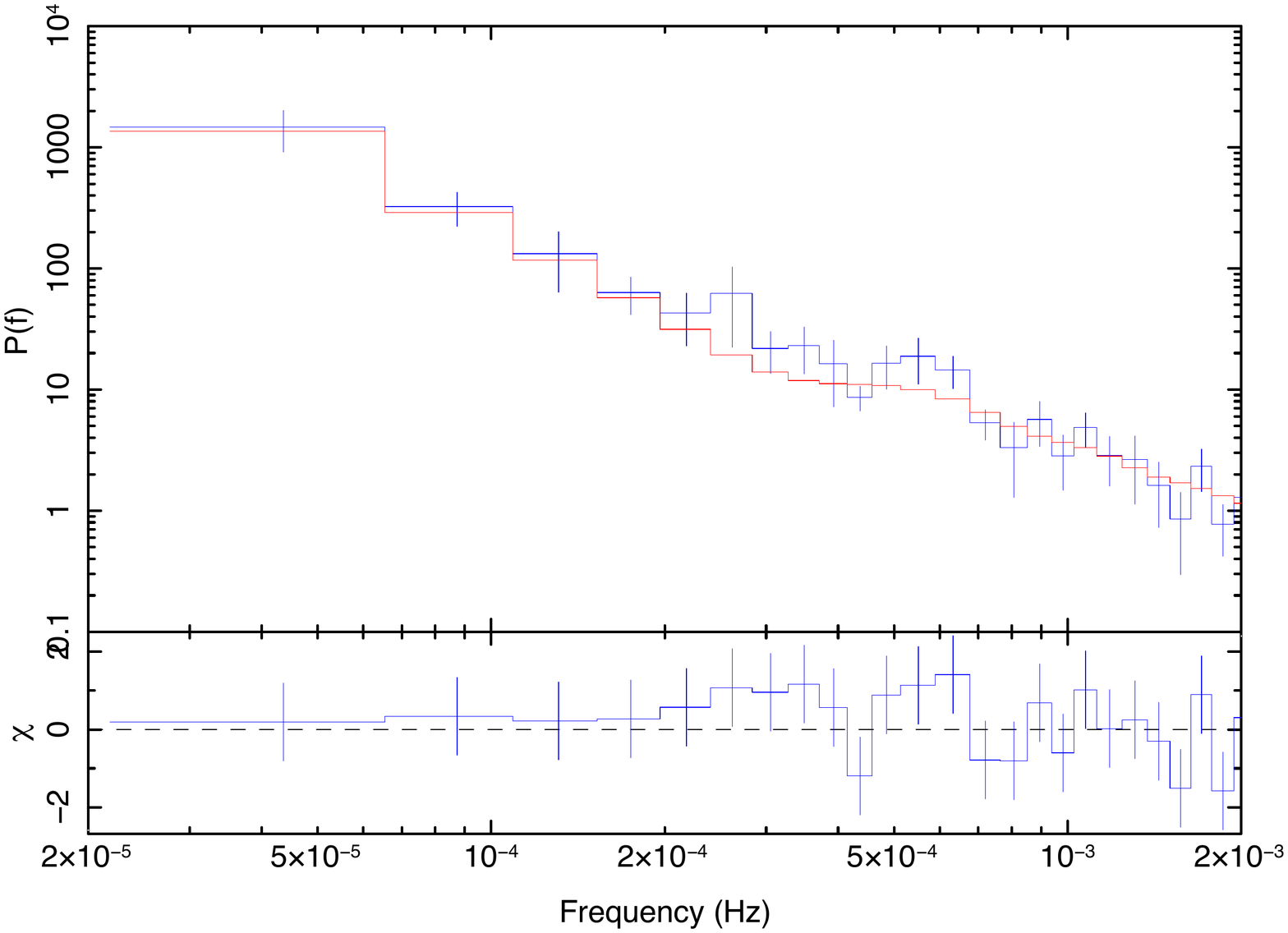}
 \put(-100,175){Rev. no. 3053}
  \put(-100,165){0.3--1 keV}
\hspace{-1.0cm}
\includegraphics[width=0.55\textwidth]{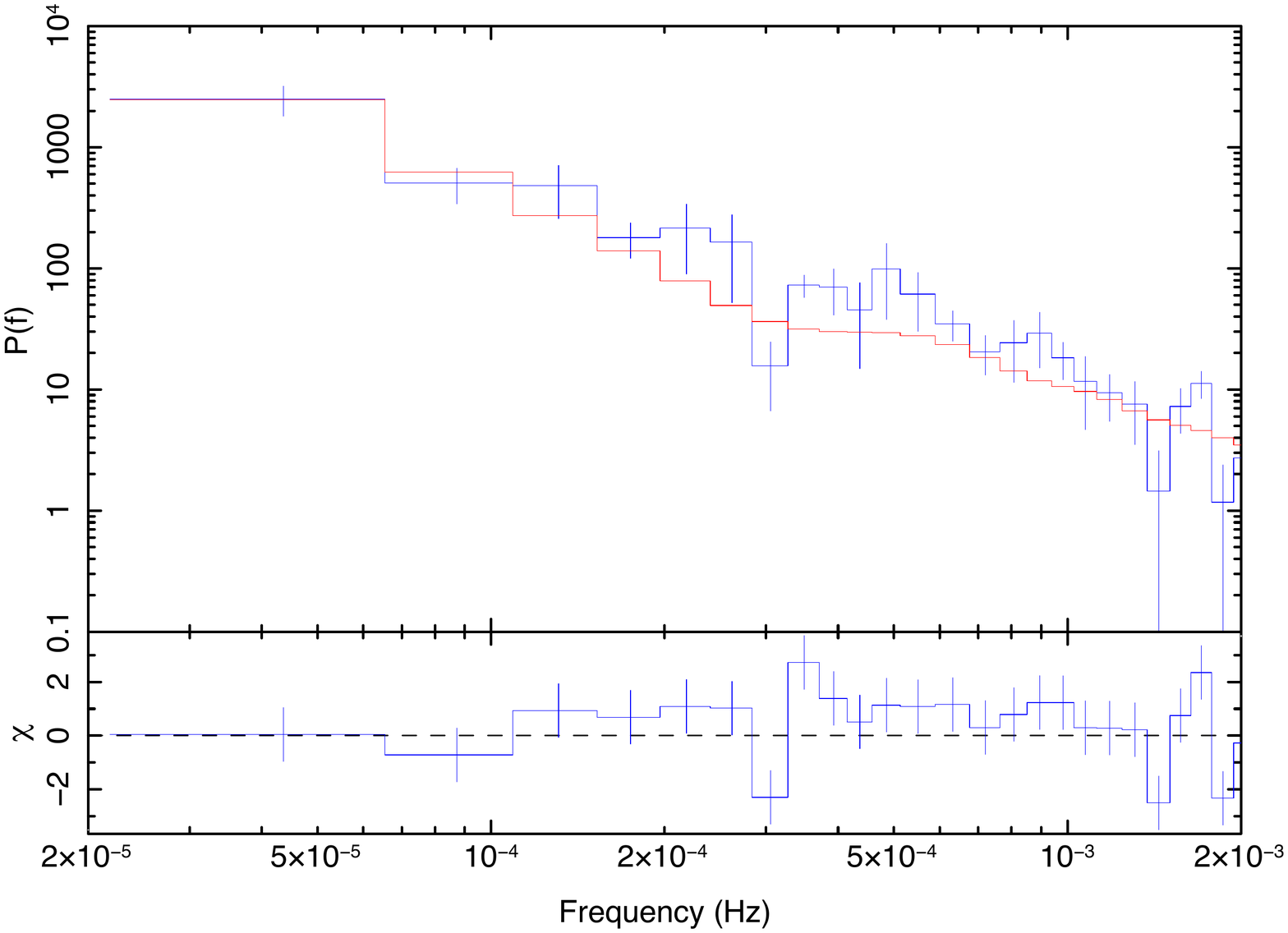}
 \put(-100,175){Rev. no. 3053}
  \put(-100,165){1.2--5 keV}
}
\caption{Data, model and residuals from simultaneously fitting the $P_{\rm rev}$ model to the data in the soft and hard bands (continued). }
\label{fig_best_fit}
\end{figure*}

\begin{table*}
\begin{center}
 \caption{Best-fit parameters from the $P_{\rm rev}$ model. The errors show 90\% confidence intervals ($\Delta \chi^2 = 2.71$). The uncertainty with ‘$-$’ is quoted when the upper or lower limit cannot be estimated due to the finite model-grid extension. 
 } 
 \label{tab_fit_para}
\begin{tabular}{ccccccccccc}
\hline
Rev. no. & $h$ & $s_{\rm s}$ & $f_{\rm b,s}$ & $R_{\rm s}$ & $A_{\rm s}$ & $s_{\rm h}$ & $f_{\rm b,h}$  & $R_{\rm h}$ & $A_{\rm h}$ & $\chi^2 / \text{d.o.f.}$ \\
 
 & ($r_{\rm g}$) &  & ($10^{-4}$ Hz) &  &  &  & ($10^{-4}$ Hz)  &  &  &  \\
\hline

2126 & $22.3^{+2.7}_{-4.5}$ & $1.38^{+0.28}_{-0.20}$ & $9.0^{+-}_{-7.2}$ & $0.56^{+0.12}_{-0.03}$ & $10.4^{+1.3}_{-6.5}$ & $1.34^{+0.29}_{-0.38}$  & $9.9^{+-}_{-8.4}$& $0.13^{+0.06}_{--}$ & $12.1^{+1.8}_{-1.6}$ & 50/31\\

2127 & $6.2^{+0.3}_{-1.2}$ & $1.65^{+0.32}_{-0.47}$ & $3.1^{+5.9}_{-2.6}$ & $0.43^{+0.15}_{-0.06}$ & $41.4^{+13.6}_{-22.1}$ & $1.63^{+0.42}_{-0.55}$  & $9.3^{+-}_{-5.5}$& $0.29^{+0.12}_{-0.14}$ & $81.9^{+32.5}_{-18.9}$ & 39/42\\

2129 & $5.7^{+4.8}_{-2.3}$ & $2.01^{+0.13}_{-0.32}$ & $9.5^{+-}_{-4.2}$ & $0.21^{+0.09}_{-0.10}$ & $24.4^{+10.2}_{-5.8}$ & $1.07^{+0.23}_{--}$  & $9.6^{+-}_{-5.4}$& $0.17^{+0.08}_{--}$ & $64.7^{+16.3}_{-18.1}$ & 28/38\\

2131 & $24.5^{+7.5}_{-6.8}$ & $2.19^{+0.06}_{-0.17}$ & $9.0^{+-}_{-1.6}$ & $0.15^{+0.05}_{--}$ & $26.6^{+17.4}_{-3.6}$ & $1.99^{+0.09}_{-0.06}$  & $9.8^{+-}_{-1.2}$& $0.12^{+0.03}_{--}$ & $47.3^{+6.0}_{-5.4}$ & 30/48\\

3037 & $14.7^{+0.9}_{-3.2}$  & $1.74^{+0.12}_{-0.10}$ & $9.8^{+-}_{-3.8}$ & $0.32^{+0.04}_{-0.07}$ & $14.6^{+1.6}_{-1.2}$ & $1.03^{+0.14}_{--}$  & $8.2^{+-}_{-1.6}$ & $0.3^{+0.02}_{-0.03}$ & $85.0^{+24.6}_{-16.9}$& 55/36\\

3038 & $12.1^{+2.7}_{-2.4}$  & $2.02^{+0.24}_{-0.32}$ & $9.2^{+-}_{-5.4}$ & $0.32^{+0.06}_{-0.10}$ & $17.4^{+11.4}_{-8.9}$ & $1.53^{+0.30}_{-0.16}$  & $9.0^{+-}_{-3.0}$ & $0.11^{+0.12}_{-0.05}$ & $26.3^{+2.3}_{-1.9}$& 37/42\\

3039 & $19.3^{+10.7}_{-8.5}$  & $2.43^{+0.04}_{-0.30}$ & $9.9^{+-}_{-3.5}$ & $0.11^{+0.12}_{--}$ & $32.3^{+3.9}_{-8.0}$ & $1.81^{+0.25}_{-0.11}$  & $9.8^{+-}_{-2.3}$ & $0.10^{+0.06}_{--}$ & $59.9^{+4.8}_{-5.4}$& 44/46\\

3043 & $8.5^{+3.0}_{-2.4}$  & $1.87^{+0.06}_{-0.26}$ & $10.0^{+-}_{-4.0}$ & $0.30^{+0.03}_{-0.05}$ & $22.6^{+2.0}_{-1.9}$ & $1.66^{+0.07}_{-0.05}$  & $9.7^{+-}_{-2.2}$ & $0.14^{+0.03}_{--}$& $42.9^{+4.6}_{-4.8}$& 52/38\\

3044 & $15.5^{+1.3}_{-0.5}$  & $1.72^{+0.19}_{-0.10}$ & $5.6^{+0.5}_{-1.1}$ & $0.37^{+0.06}_{-0.02}$ & $15.0^{+1.9}_{-2.2}$ & $1.69^{+0.06}_{-0.03}$  & $7.1^{+0.4}_{-0.8}$ & $0.26^{+0.02}_{-0.01}$& $55.9^{+2.2}_{-1.9}$& 78/48\\

3045 & $14.8^{+1.2}_{-2.2}$  & $2.43^{+0.07}_{-0.18}$ & $9.5^{+0.3}_{-1.8}$ & $0.15^{+0.04}_{-0.04}$ & $33.4^{+4.6}_{-5.5}$ & $1.83^{+0.13}_{-0.20}$  & $9.8^{+-}_{-1.5}$ & $0.14^{+0.02}_{-0.03}$& $44.1^{+3.2}_{-3.3}$& 45/46\\

3046 & $10.2^{+2.3}_{-2.6}$  & $1.96^{+0.05}_{-0.05}$ & $9.2^{+0.6}_{-1.2}$ & $0.29^{+0.05}_{-0.03}$ & $49.7^{+15.1}_{-5.4}$ & $1.77^{+0.12}_{-0.09}$  & $9.9^{+-}_{-1.7}$ & $0.14^{+0.04}_{-0.05}$& $94.2^{+7.7}_{-6.3}$& 35/39\\

3048 & $10.8^{+1.3}_{-1.1}$  & $2.00^{+0.29}_{-0.34}$ & $1.0^{+2.3}_{--}$ & $0.30^{+0.21}_{-0.06}$ & $71.6^{+39.5}_{-41.7}$ & $2.06^{+0.05}_{-0.04}$  & $9.8^{+-}_{-2.8}$ & $0.10^{+0.03}_{-0.05}$& $100.5^{+19.4}_{-18.8}$& 46/44\\

3049 & $3.0^{+0.8}_{--}$  & $1.81^{+0.28}_{-0.19}$ & $2.0^{+2.7}_{-1.1}$ & $0.34^{+0.09}_{-0.11}$ & $65.9^{+7.9}_{-11.8}$ & $1.61^{+0.14}_{-0.06}$  & $9.5^{+0.3}_{-1.6}$ & $0.25^{+0.03}_{-0.04}$& $100.0^{+28.7}_{-14.6}$& 52/47\\

3050 & $11.9^{+2.7}_{-3.0}$  & $1.85^{+0.08}_{-0.19}$ & $8.6^{+1.2}_{-2.1}$ & $0.30^{+0.03}_{-0.08}$ & $11.8^{+2.0}_{-1.9}$ & $1.68^{+0.06}_{-0.07}$  & $9.7^{+-}_{-3.7}$ & $0.25^{+0.06}_{-0.02}$& $27.8^{+3.5}_{-3.7}$& 84/47\\

3052 & $19.4^{+12.6}_{-6.5}$  & $1.86^{+0.24}_{-0.11}$ & $3.0^{+4.8}_{-1.2}$ & $0.11^{+0.14}_{--}$ & $12.7^{+4.3}_{-1.5}$ & $1.76^{+0.31}_{-0.06}$  & $9.6^{+-}_{-4.9}$ & $0.11^{+0.10}_{--}$& $18.7^{+2.1}_{-2.3}$& 54/47\\

3053 & $11.6^{+2.2}_{-0.8}$  & $1.61^{+0.21}_{-0.05}$ & $1.4^{+2.8}_{-0.3}$ & $0.33^{+0.03}_{-0.18}$ & $11.6^{+4.2}_{-3.0}$ & $1.58^{+0.05}_{-0.04}$  & $6.8^{+3.0}_{-2.5}$ & $0.30^{+0.04}_{-0.1}$& $24.9^{+2.4}_{-2.5}$& 72/49\\

\hline
\end{tabular}
\end{center}
\end{table*}
\nopagebreak

\begin{figure}
    \centering
    \includegraphics[width=0.47\textwidth]{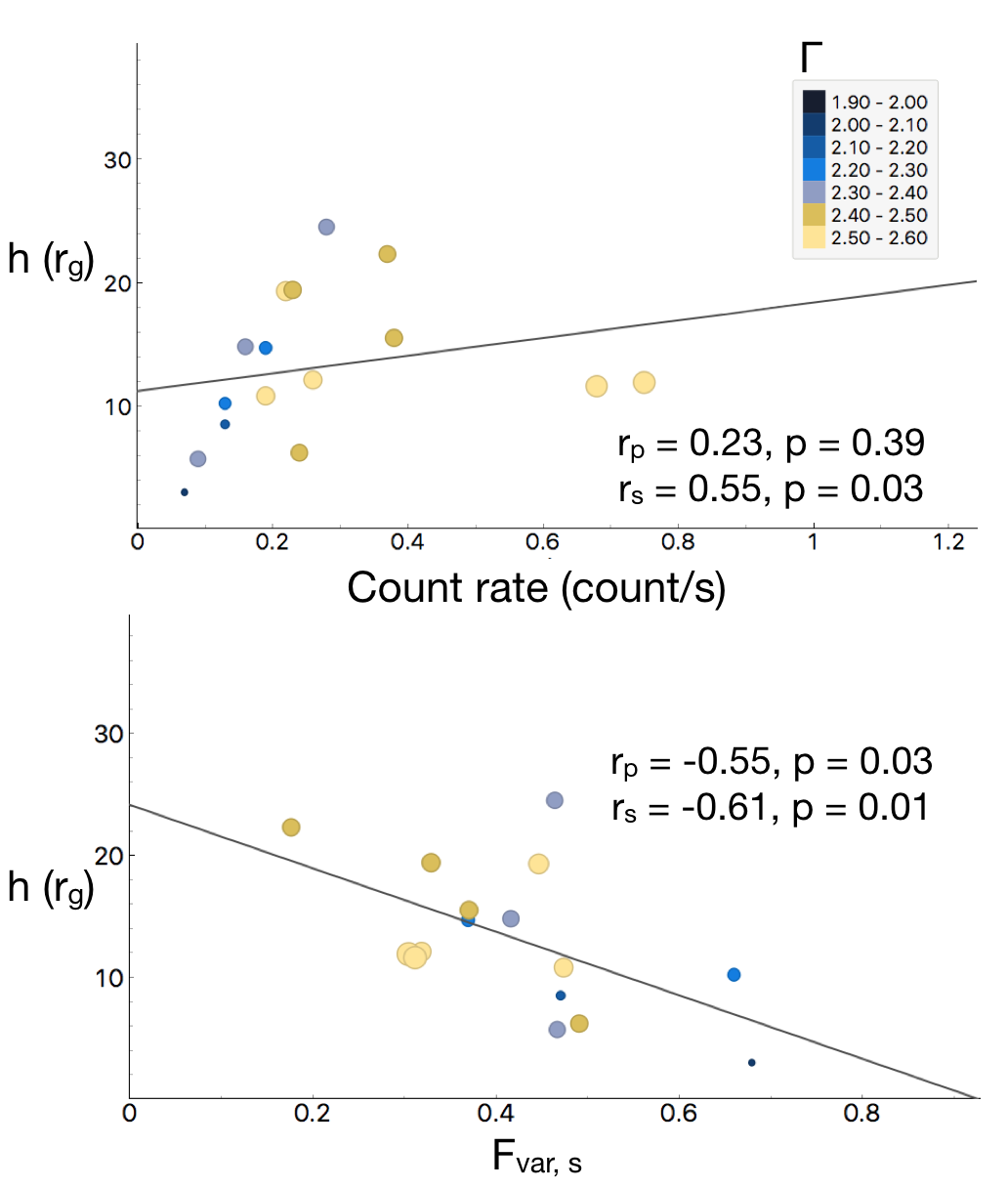}
    \caption{Source height ($h$) versus count rate (top panel) and soft fractional excess variance ($F_{\rm var, s}$) (bottom panel). The data size and color represent the photon index ($\Gamma$) from the spectral fits of \cite{Alston2020}. The solid lines represent the regression lines, with the Pearson and Spearman correlation coefficients and $p$-values illustrated. }
    \label{fig-h}
\end{figure}

\begin{figure*}
    \centering
    \includegraphics*[scale=0.55]{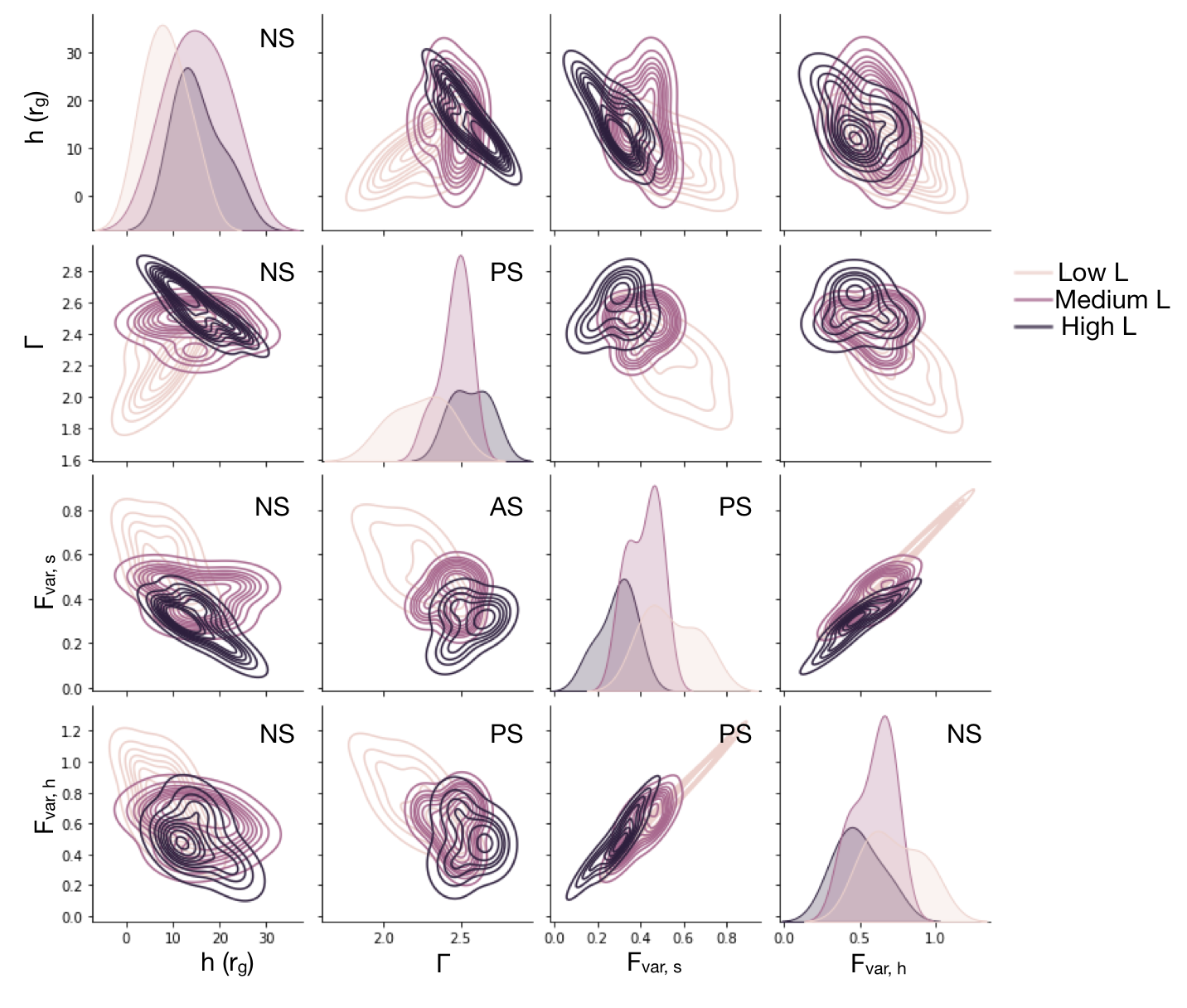}
    \caption{Pairwise relationships between the variable $h$, $\Gamma$, $F_{\rm var, s}$ and $F_{\rm var, h}$, with the data being divided into 3 groups: low, medium and high luminosity as noted in Table~\ref{tab_obs}. The distributions are derived using 16 observations of IRAS~13224--3809, with a layered kernel density estimate (KDE). The diagonal plots represent a univariate marginal distribution of the data in each column. The labels AS, PS, and NS mean that the visual differences of all 3 grouped samples are significant, only pair of them is significant, and none of them is significant, respectively. See text for more details.  
     }
    \label{fig-seaborn}
\end{figure*}

Firstly, we used the reverberation PSD model ($P_{\rm rev}$) to fit the individual PSD data of IRAS~13224--3809 without including the Lorentzian component. The reverberation effect imprinted in the PSD profiles can plausibly explain the oscillatory features seen in the data. Examples of the fitting results for certain individual observations are presented in Fig.~\ref{fig_best_fit}. Some observations show clues of narrow features, but due to the poor quality of the data (low signal-to-noise), these narrow features cannot be robustly identified. In fact, even without the additional Lorentzian component, the model can provide good fits for the majority of the data, with adequate fits in some observations. We then select to exclude the Lorentzian component when we analyze the individual observations, and include it later when the data are combined using the luminosity bin.

The best-fit parameters for 16 observations of IRAS~13224--3809 are presented in Table~\ref{tab_fit_para}. Note that, for each observation, the PSD profiles extracted in the soft and hard bands are simultaneously fitted, with the source height ($h$) being tied together. The errors correspond to 90\% confidence intervals ($\Delta \chi^2 = 2.71$) around the best-fitting values and, where necessary, are estimated by linear interpolation between the consecutive model-grid spacing for each parameter. The obtained reflection fraction in the soft band is larger than those in the hard band, which is consistent with the X-ray reflection framework, in which the reflection flux is more dominant in the soft band. We find the average values of $f_{\rm b, s} \sim 6.9 \times 10^4$~Hz, $f_{\rm b, h} \sim 9.2 \times 10^4$~Hz, $s_{\rm s} \sim 1.91$ and $s_{\rm h} \sim 1.63$. The source height found in these 16 observations varies between $h \sim 3$--$25 \ r_{\rm g}$.

In Fig.~\ref{fig-h}, the source height is plotted against the count rate, as well as the fractional excess variance observed in the soft band ($F_{\rm var, s}$). To elaborate more on the parameter correlations, we employ the photon index ($\Gamma$) from the average spectral model fits of \cite{Alston2020}. The $\Gamma$ values are represented by the size and color of the data points. The data correlations and visualizations are analyzed using the Orange data mining platform in Python \citep{Demsar2013}. We find that the Pearson and Spearman correlation coefficients between $h$ and count rate are $r_{\rm p} = 0.23$ and $r_{\rm s} =0.55$, with the $p$-values of 0.39 and 0.03, respectively. This suggests that monotonic relationships between $h$ and count rate are not statistically significant ($p > 0.01$). This is probably due to the small sample and large scatter. On the other hand, the variables $h$ and $F_{\rm var, s}$ are moderately anti-correlated in a non-linear way, with $r_{\rm s} =-0.61$ ($p=0.01$).



Fig.~\ref{fig-seaborn} represents the distribution of the parameters $h$, $\Gamma$, $F_{\rm var, s}$ and $F_{\rm var, h}$ where the observational data are divided into three different luminosity groups (i.e., low, medium, and high luminosity), as specified in Table~\ref{tab_obs}. A layered kernel density estimate (KDE) is used to create the sample density distribution for each group. The observations displaying larger $L$ seem to relate with larger source $h$ and $\Gamma$, and smaller $F_{\rm var}$. Obviously, $F_{\rm var, s}$ is correlated with $F_{\rm var, h}$. Furthermore, we perform both univariate and multivariate nonparametric two-sample test with bootstrap probabilities using Cramer-Test implemented in CRAN package \citep{Feigelson2012} in order to verify if the visual differences of the plots between 3 luminosity groups are significant. We produce 1,000 bootstrap-replicates with the normal Monte-Carlo-bootstrap method and set the confidence level of test to be 95\%. Our hypothesis is that the samples in one luminosity group can be distributed as those in another group, which is accepted or rejected based on the estimated $p$-value. We find that the sample distributions in the $F_{\rm var,s}$--$\Gamma$ space are clearly and significantly separated. The patterns of associations between $h$ and $F_{\rm var}$ in both energy bands also show a hint of separation of sources in different luminosity bins, but the differences are not significant.   

\begin{figure}
    \centering
    \includegraphics[width=0.47\textwidth]{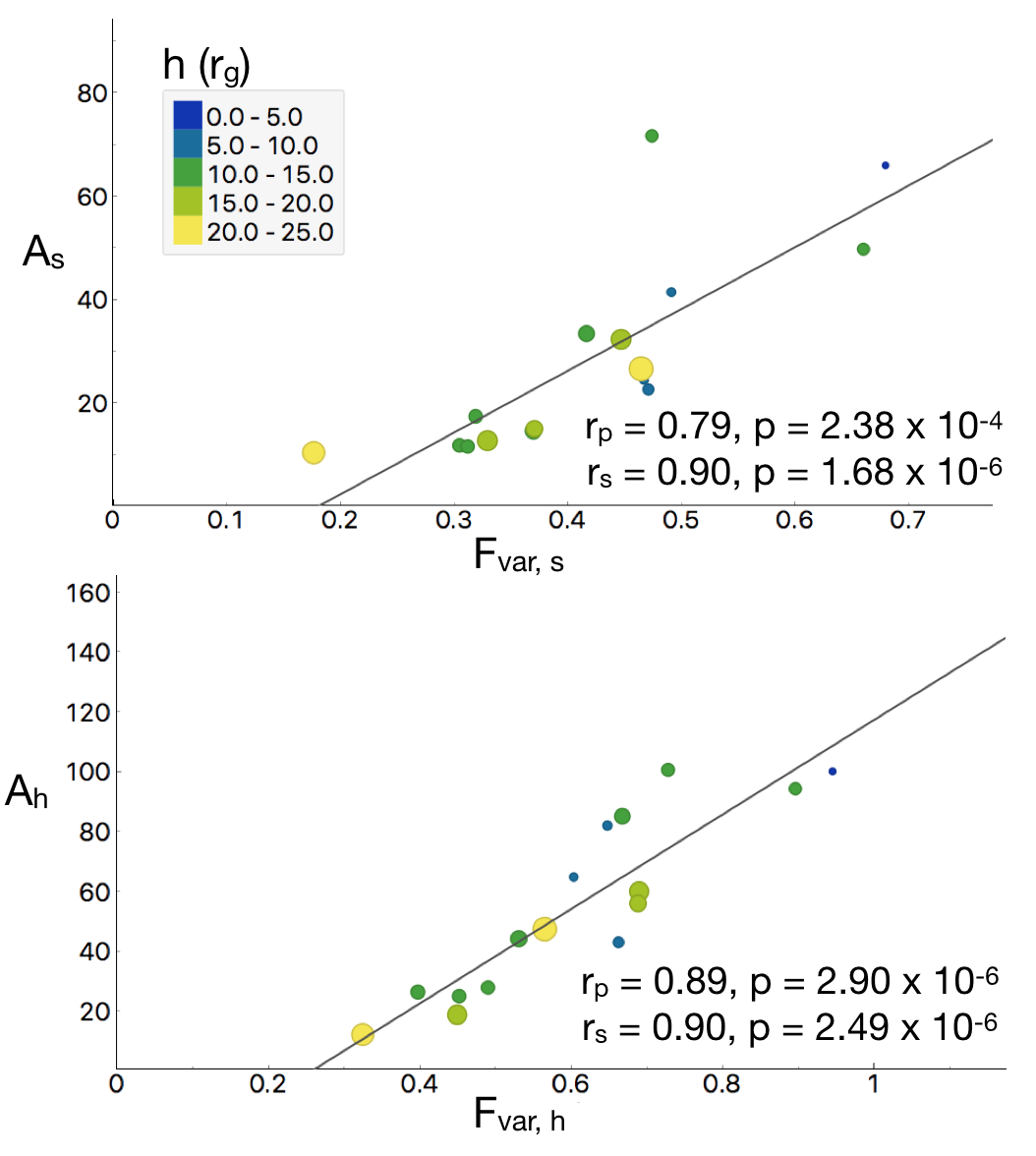}
    \caption{Normalization factor ($A$) versus the fractional excess variance ($F_{\rm var}$) for the soft (top panel) and the hard (bottom panel) bands, where the size and color of the data points show the obtained source height ($h$). The solid lines represent the regression lines, with the Pearson and Spearman correlation coefficients shown.}
    \label{fig-norm}
\end{figure}

\begin{figure*}
    \centering
    \includegraphics*[scale=0.6]{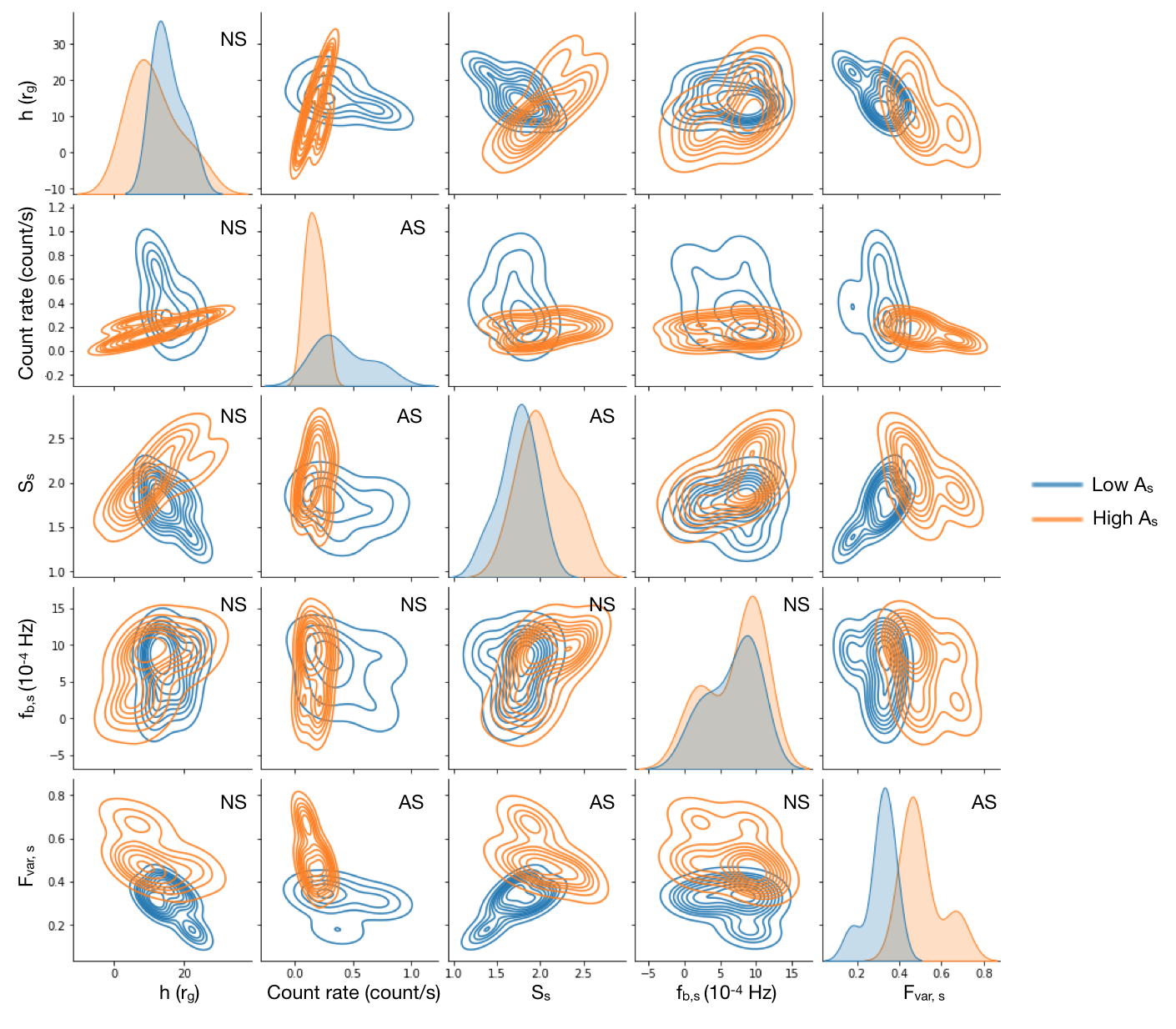}
    \caption{Pairwise relationships between the obtained source height ($h$), count rate, bending index ($s_{\rm s}$), bending frequency ($f_{\rm b, s}$) and fractional excess variance ($F_{\rm var, s}$). The density distribution is produced via KDE, with two PSD normalization intervals: low and high $A_{\rm s}$ of $<20$ and $\geq 20$, respectively. The labels AS and NS mean that the visual differences are and are not significant, respectively}. 
    \label{fig-As}
\end{figure*}

Fig.~\ref{fig-norm} shows the correlations between the constrained $A$ and $F_{\rm var}$ for both energy bands. We observe a high correlation between $A$ and $F_{\rm var}$, with $r_{\rm p}= 0.79$ and 0.89 for the soft and hard bands, respectively, and $r_{\rm s}= 0.90$ for both energy bands. Furthermore, Fig.~\ref{fig-As} shows the pairwise relationships between the PSD parameters of the soft band with the KDE density distribution, where the sources are divided into two groups: low and high $A_{\rm s}$. The multivariate two-sample test with bootstrap probabilities, as mentioned earlier, is also performed to identify whether the visual differences are significant. We can see clearly that the group of low $A_{s}$ most likely corresponds to a source with smaller $s_{\rm s}$ and $F_{\rm var, s}$ (blue profiles along the diagonal plots significantly shift to the left). However, the association of $A_{\rm s}$ with $f_{\rm b, s}$ cannot be clearly resolved (i.e. the base line of blue and orange profiles along the diagonal plots almost covers the same parameter space).

\begin{figure}
    \centering
    \includegraphics*[scale=0.50]{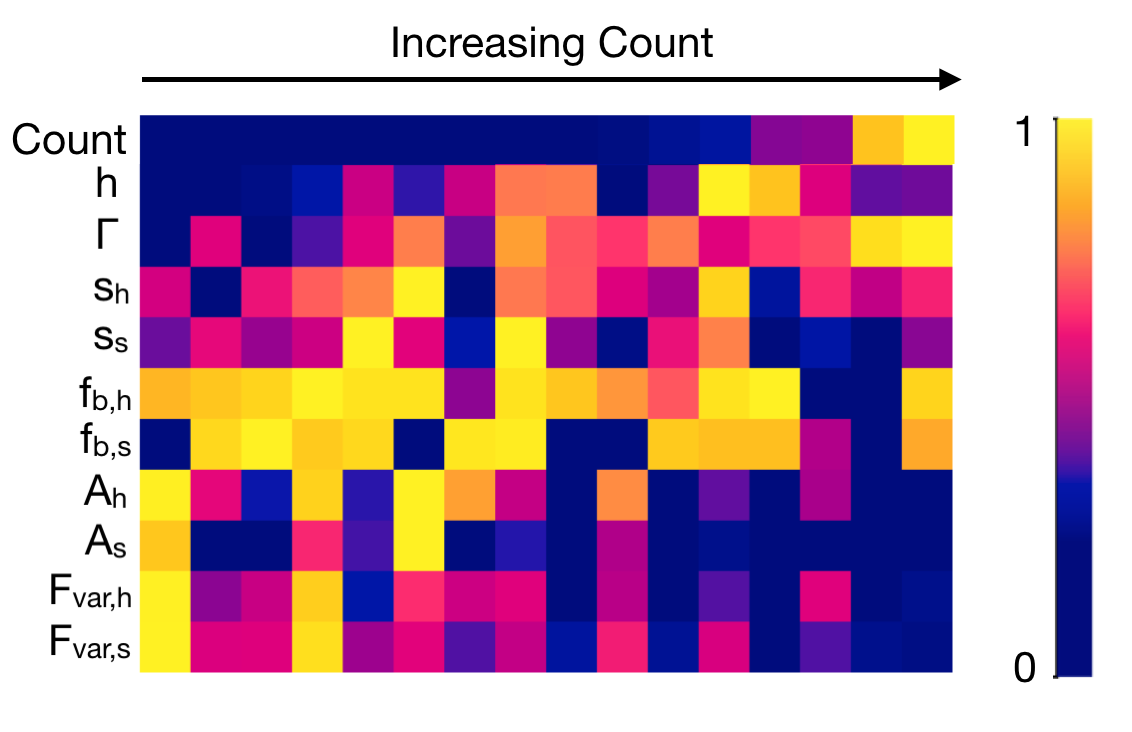}
    \caption{Heat map showing how the overall parameters change with increasing count rate. The values for each parameter are normalized in the range of 0 and 1 in order to highlight the parameter evolution.}
    \label{fig-params-dis}
\end{figure}

\begin{figure}
    \centering
    \includegraphics*[scale=0.60]{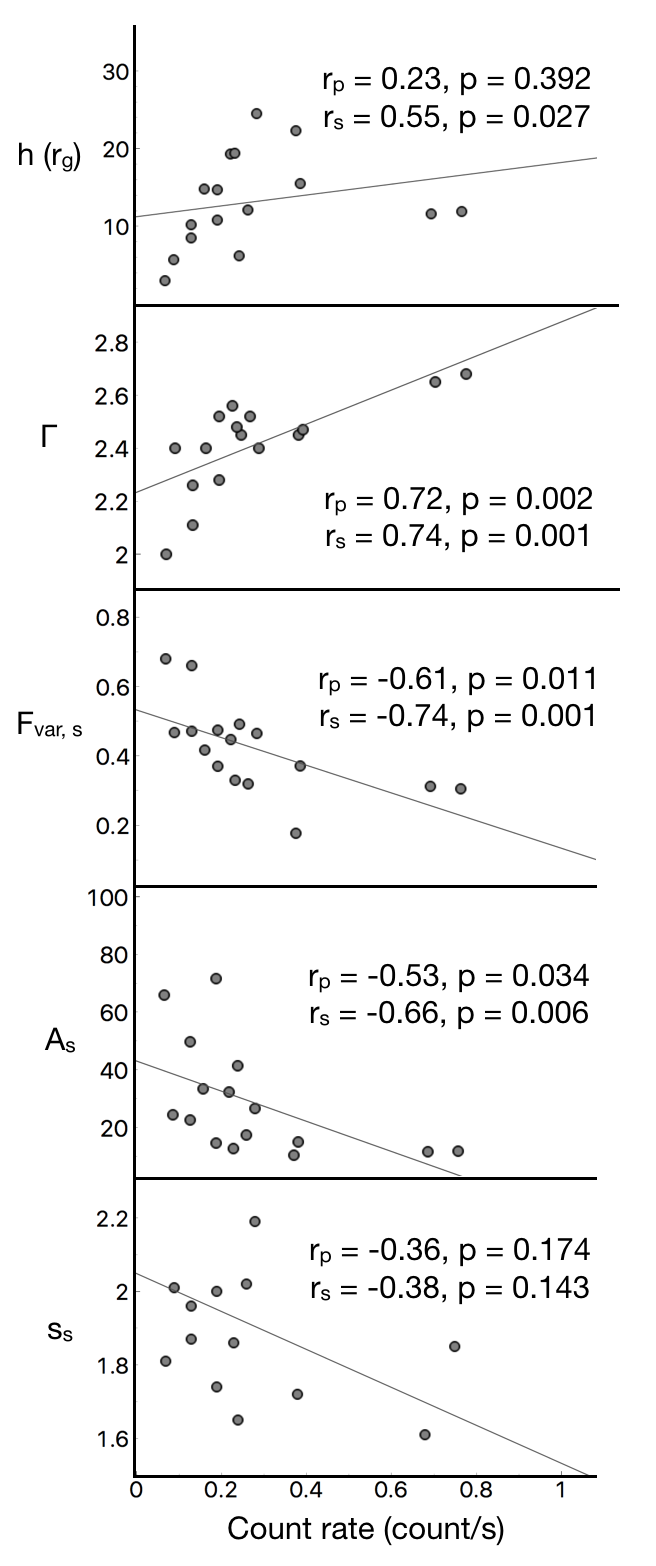}
    \caption{Scatter plots showing the overall correlations of the model parameters. All are plotted against the count rate (i.e., source luminosity). }
    \label{fig-sum}
\end{figure}

Furthermore, we illustrate how the overall parameters of IRAS~13224--3809 change as a heat map in Fig.~\ref{fig-params-dis}, where the obtained values for each parameter are transformed to a scale from 0 to 1. This clearly reveals that the observation with higher count rate seems to have higher $h$ and $\Gamma$, but smaller $s$, $f_{b}$, $A$, and $F_{\rm var}$. This implies an increase of source height occurring with a decrease of the observed $F_{\rm var}$, with the flatter PSD slope. Scatter plots representing the overall correlations of the model parameters with respect to the count rate are also shown in Fig.~\ref{fig-sum}. A hint of moderate-to-strong monotonic relationships between the count rate and $h$, $\Gamma$, $F_{\rm var,s}$, and $A_s$ ($\mid r_{\rm s} \mid > 0.5 $) can be observed.

Due to the small sample, we also examine the distribution of correlation coefficients with paired-bootstrap resampling in order to construct the 95\% confidence interval of the paired-sample statistics. This is to ensure that the correlation coefficients that show $p < 0.01$ in Fig.~\ref{fig-sum} are certain. The random sample is taken with replacement from our 16 samples to form a paired-bootstrap sample similar size to the original sample. This process is repeated 1,000 times to produce the paired-bootstrap distribution. For the count rate and $\Gamma$, we find that $0.34 \leq r_{\rm s} \leq 0.93$, with 95\% confidence. We also find that $-0.93 \leq r_{\rm s} \leq -0.38$ for the count rate and $F_{\rm var,s}$, and $-0.85 \leq r_{\rm s} \leq -0.28$ for the count rate and $A_{\rm s}$. The correlation coefficients then seem to be varied, but the trend in which each parameter is correlated or anti-correlated with count rate is quite certain despite of our limited sample. 

\begin{figure*}
\centerline{
\includegraphics*[width=0.55\textwidth]{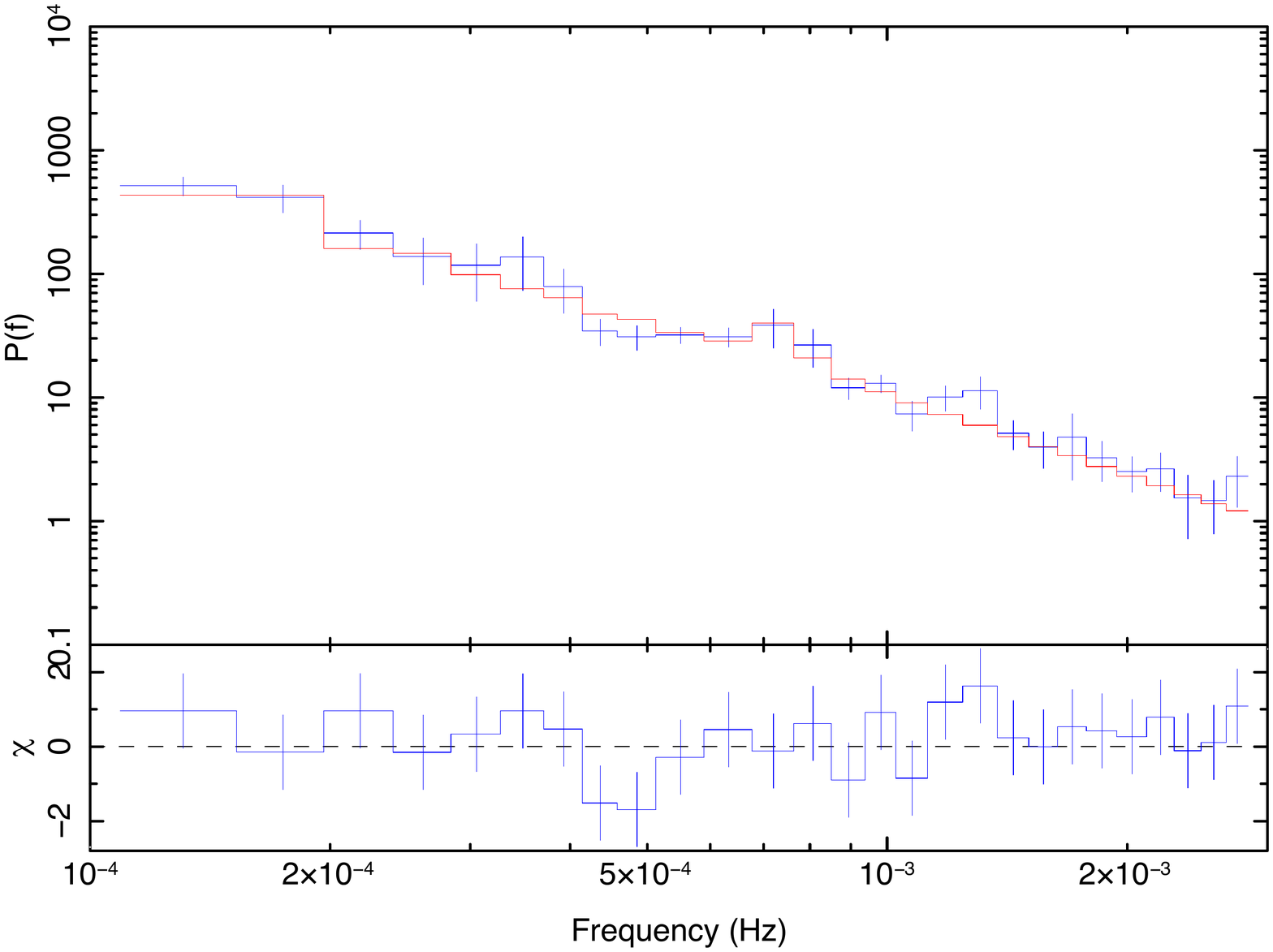}
 \put(-110,175){Low luminosity}
  \put(-110,165){0.3--1 keV}
\hspace{-1.0cm}
\includegraphics*[width=0.55\textwidth]{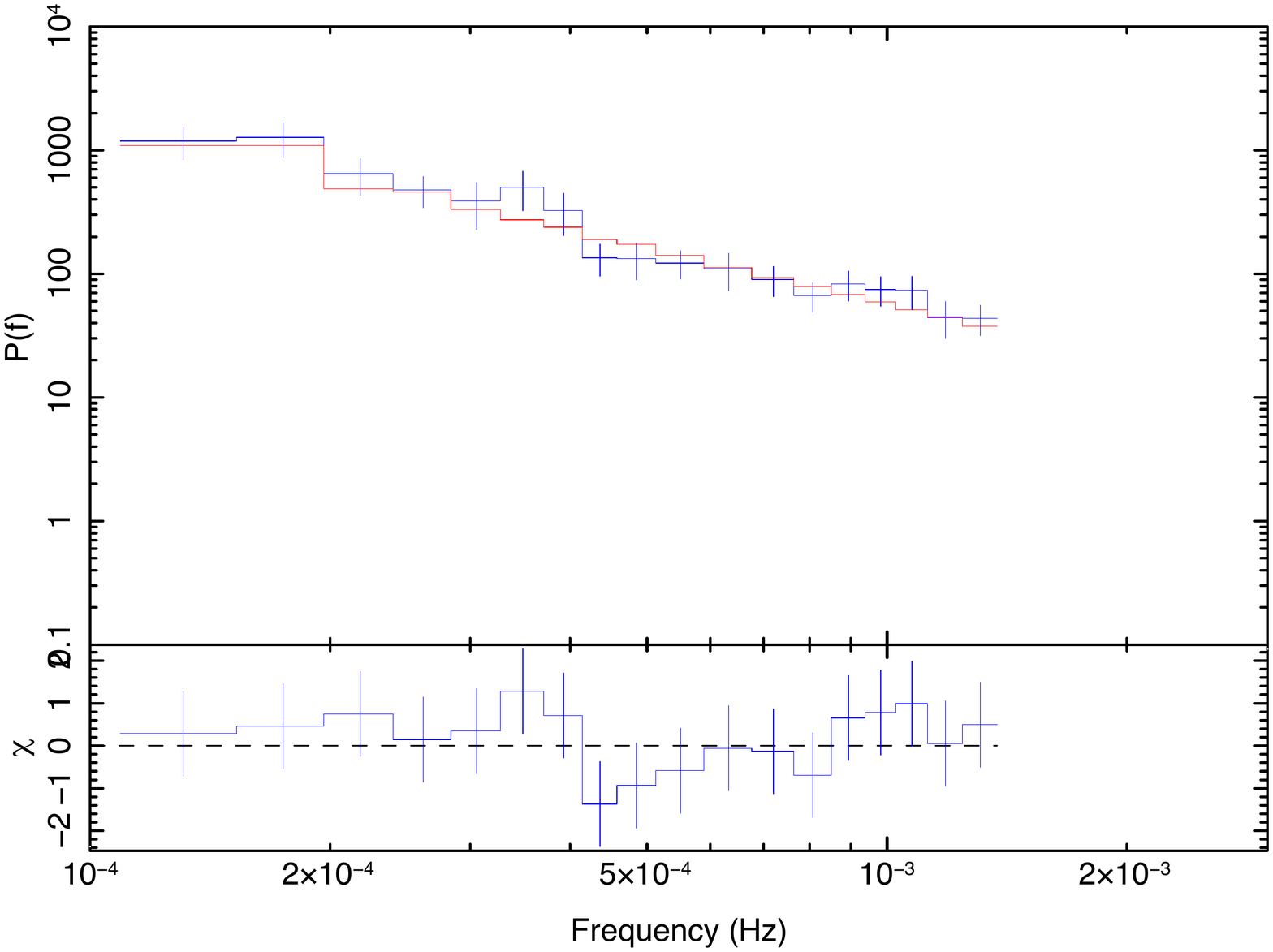}
 \put(-110,175){Low luminosity}
  \put(-110,165){1.2--5 keV}
\vspace{-0.5cm}
}
\centerline{
\includegraphics[width=0.55\textwidth]{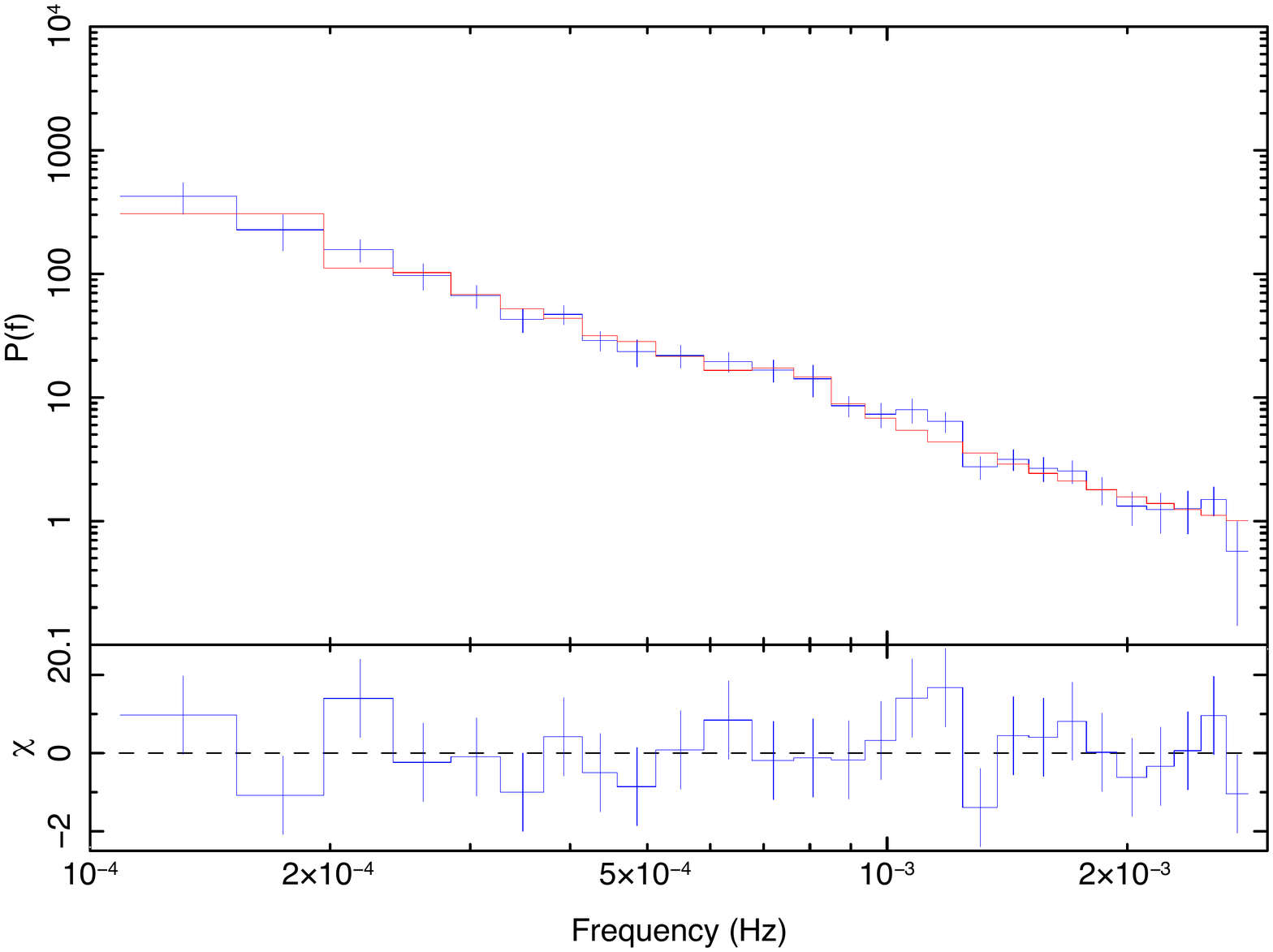}
 \put(-110,175){Medium luminosity}
  \put(-110,165){0.3--1 keV}
\hspace{-1.0cm}
\includegraphics[width=0.55\textwidth]{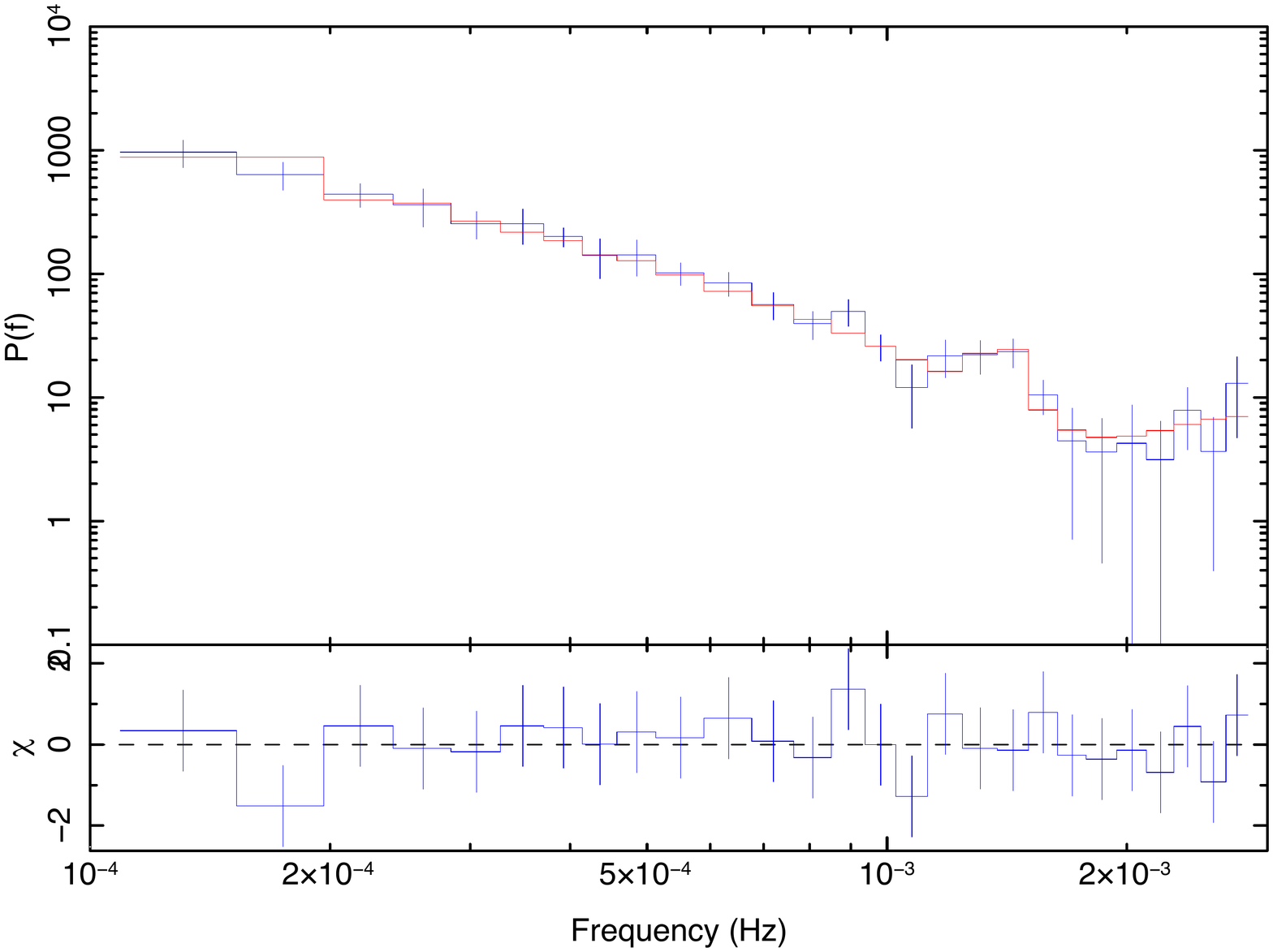}
 \put(-110,175){Medium luminosity}
  \put(-110,165){1.2--5 keV}
\vspace{-0.5cm}
}
\centerline{
\includegraphics[width=0.55\textwidth]{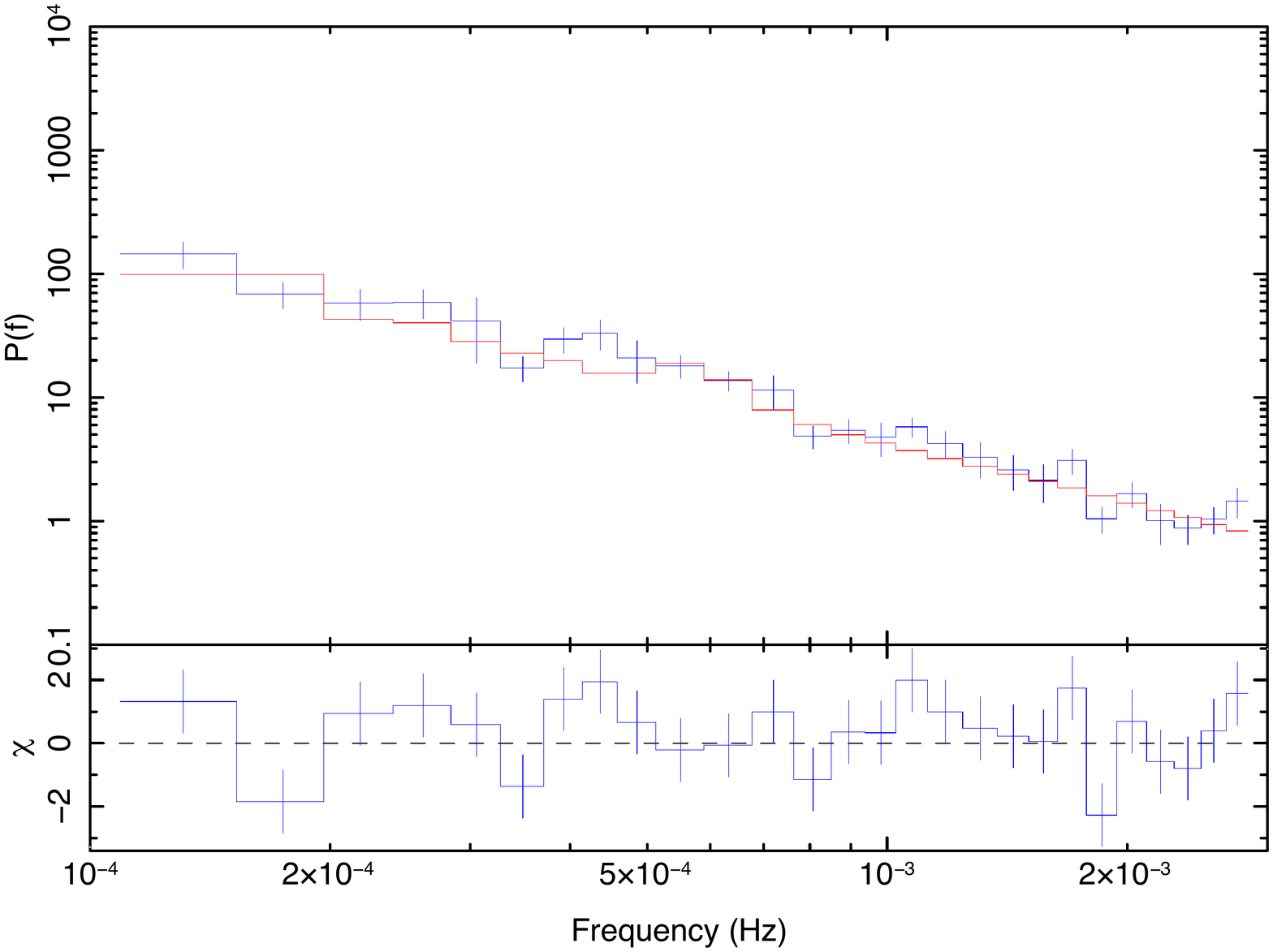}
 \put(-110,175){High luminosity}
  \put(-110,165){0.3--1 keV}
\hspace{-1.0cm}
\includegraphics[width=0.55\textwidth]{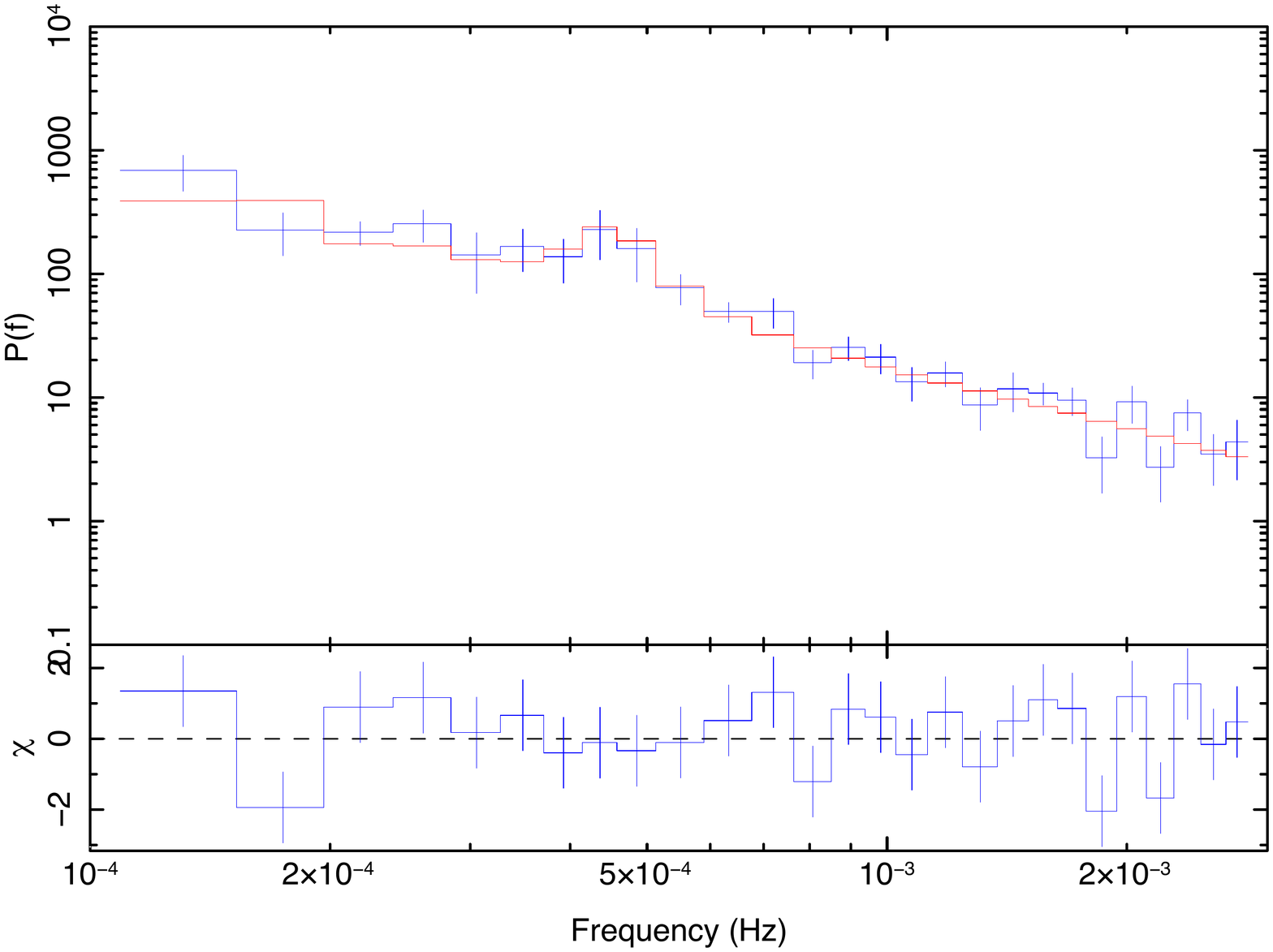}
 \put(-110,175){High luminosity}
  \put(-110,165){1.2--5 keV}
}
\caption{Data, model and residuals from fitting the ($P_{\rm rev}+P_{\rm lor}$) model (red) to the data (blue) extracted in the 0.3--1~keV band (left panels) and 1.2--5~keV band (right panels) for different luminosity bins: low (top panels), medium (middle panels) and high (lower panels) luminosity. The obtained parameters are listed in Table~\ref{tab_fit_para_group}. }
\label{fig_best_fit_group}
\end{figure*}

Now, we fit the PSD of IRAS~13224--3809 when the observational data are combined into three groups: low, medium and high luminosity. In this case the signal-to-noise ratio is relatively high compared to when we fit each individual PSD data. Therefore, the additional Lorentzian component is included when necessary to explain the narrow features appearing in the profiles. The line's centroid frequency ($f_{\rm lor}$) and FWHM ($\sigma_{\rm lor}$) are allowed to be free. The best-fit results are presented in Fig.~\ref{fig_best_fit_group} and the parameters are shown in Table~\ref{tab_fit_para_group}. The model can provide good fits for the low and medium luminosity spectra, with adequate fits for the high luminosity spectrum. Both energy bands show a hint of the narrow peak moving towards lower frequencies as the source luminosity increases, especially when considering the medium and high luminosity bins. The FWHM of the narrow line is $\sigma_{\rm lor} \sim 10^{-4}$~Hz for both energy bands and all luminosity bin spectra, with the exception of the hard band of low luminosity data where the narrow feature is not significantly required. 

The results from the luminosity bin spectra also show that the PSD shape flattens as energy increases ($s_{\rm s} > s_{\rm h}$), but there is a less clear evolution of the bending PSD index with luminosity. Despite of this, the results still support the trend of increasing source height with increasing luminosity, from $\sim 3 \ r_{\rm g}$ to $\gtrsim 20 \ r_{\rm g}$, as predicted by the individual fitting.

\begin{table*}
\begin{center}
\caption{Best-fit parameter values for the ($P_{\rm rev}+P_{\rm lor}$) model. The first and second columns show the grouped data according to the luminosity bin and the source height ($h$), respectively. Note that $s$ and $f_{\rm b}$ is the bending index and bending frequency of the PSD, respectively, while $f_{\rm lor}$ and $\sigma_{\rm lor}$ is the centroid frequency and the FWHM of the Lorentzian line profile. The $\chi^2 / \text{d.o.f.}$ for the low, medium and high luminosity bins are of 27/33, 30/39, and 63/39, respectively.}
 \label{tab_fit_para_group}
\begin{tabular}{cccccccccc}
\hline
Luminosity & $h$ & $s_{\rm s}$ & $f_{\rm b,s}$ & $f_{\rm lor, s}$ & $\sigma_{\rm lor, s}$ & $s_{\rm h}$ & $f_{\rm b,h}$  & $f_{\rm lor, h}$ & $\sigma_{\rm lor, h}$  \\
 
bin & ($r_{\rm g}$) &  & ($10^{-4}$ Hz) & ($10^{-4}$ Hz) & ($10^{-4}$ Hz) &  & ($10^{-4}$ Hz)  & ($10^{-4}$ Hz) & ($10^{-4}$ Hz)   \\
\hline

Low & $3.3^{+2.5}_{-1.0}$ & $1.78^{+0.18}_{-0.12}$ & $1.2^{+0.4}_{-0.2}$ & $7.3^{+0.1}_{-0.2}$ & $1.0^{+1.2}_{-2.5}$ & $1.37^{+0.22}_{-0.18}$  & $1.2^{+0.3}_{-0.2}$& $-$ & $-$ \\

Medium & $10.6^{+2.8}_{-1.6}$ & $1.74^{+0.14}_{-0.10}$ & $1.1^{+0.2}_{-0.1}$ & $7.6^{+0.1}_{-0.3}$ & $1.0^{+1.0}_{-2.4}$ & $1.38^{+0.16}_{-0.12}$ & $9.6^{+-}_{-1.6}$  & $13.8^{+0.2}_{-0.3}$ & $1.1^{+1.1}_{-1.8}$  \\

High & $22.4^{+7.2}_{-4.2}$ & $1.74^{+0.12}_{-0.12}$ & $9.5^{+-}_{-0.2}$ & $5.8^{+0.1}_{-0.2}$ & $1.0^{+1.0}_{-1.2}$ & $1.67^{+0.15}_{-0.12}$ & $9.7^{+-}_{-0.6}$  & $4.5^{+0.4}_{-0.5}$ & $1.0^{+1.2}_{-1.0}$  \\

\hline
\end{tabular}
\end{center}
\end{table*}
\nopagebreak

\section{Discussion}

We investigate the corona evolution of IRAS~13224--3809 using 16 {\it XMM-Newton} observations solely through the PSD analysis. We take into account the reverberation effects caused by the lamp-post source, which represent an echo filter to the intrinsic variability signals  \citep[e.g.][]{Papadakis2016, Chainakun2019a}. There might be the systematic uncertainties due to the choices of $\Gamma$ and $i$. While \cite{Alston2020} derived $\Gamma$ from the reflection fits using the low-density disk, \cite{Jiang2022} considered a high-density model and found $\Gamma$ is consistently higher. Note that $\Gamma$ could change the flux contribution in each energy band, which should be compensated with allowing the reflection fraction to be free. \cite{Jiang2022} also reported $i \sim 60^\circ$--$70^{\circ}$, while in this work we use the intermediate value of $i = 45^{\circ}$ as in \cite{Caballero2020}. These uncertainties should not have significant effects in our key results. 
Nevertheless, we notice that the reverberation signatures on the PSD slightly change with inclinations \citep{Papadakis2016}.

By fitting the soft and hard PSD simultaneously, we find that the source height $h$ seems to be correlated with luminosity. This is consistent with \cite{Alston2020} and \cite{Caballero2020} who utilized different spectral and temporal profiles and reported a tendency of increasing source height with luminosity. Furthermore, \cite{Chainakun2022} investigated the correlations between the reverberating AGN parameters and reported an anticorrelation between the Fe-K lag and $F_{\rm var}$. Given that the amplitude of the lag (either in Fe-K or Fe-L band) increases with $h$, the hint of the anticorrelation between $h$ and $F_{\rm var}$ found here is well justified. This demonstrates that a consistent framework can be inferred even when diverse timing data and analyses are used.

The source height in this work is varying between $\sim 3$--$25 \ r_{\rm g}$. The spectral analysis by \cite{Jiang2022} using almost the same set of observations revealed that the source height $h$ varied between 0.43--1.71~$\sqrt{f_{\rm AD} / f_{\rm INF}}$~$r_{\rm g}$, where $f_{\rm AD} / f_{\rm INF}$ is the ratio between the coronal flux that reaches the accretion disk and the flux at infinity, which is geometry dependent. By assuming a lamp-post geometry and using the best-fit ionization and density parameters, they approximated the source location to be of $3$--$6 \ r_{\rm g}$, which is much smaller than our analysis. This provides further evidence for the inconsistency in the parameters obtained from the timing (reverberation) data and the time-averaged spectral analysis, which was recently reported in several AGN such as NGC~5506 \citep{Zoghbi2020} and MCG--5--23--16 \citep{Zoghbi2021}.

Meanwhile, the spectral components of IRAS~13224--3809 have been found to be complicated, particularly in the soft excess below 2 keV \citep[e.g.][]{Fabian2013,Chainakun2016}. The time-averaged spectrum, for example, can be fitted with either the standard disk model with an additional soft-excess blackbody component \citep{Jiang2018} or the reflection from a high density disk model \citep{Jiang2022}. Different spectral models result in different amount of dilutions applied to the reverberation lags (e.g. increasing the disk density results in an increase of reflection flux especially in the soft band of $<2$~keV). Here, we do not determine which model spectrum is preferable over the others; instead, we choose to let the reflection fraction $R$ to be free. There is also the change in flux contribution if we consider the returning radiation \citep{Wilkins2020}, or take into account the flux from the ultra-fast outflows \citep{Parker2017,Parker2021} that may be present. The parameter $R$ then takes into account the impacts of the energy band chosen as well as incorporates the spectral complexity which is not explicitly modelled here. Nevertheless, the best-fit model prefers  $R_{\rm s} \gtrsim R_{\rm h}$, which is consistent with the reflection framework that the reflection flux is mostly contributing in the soft band.

The intrinsic shape of the PSD represents the variability power that tightly depends on the physical properties of the accretion flow in the framework of the propagation of mass
accretion \citep[e.g.][]{Arevalo2006, Ingram2011,Mahmoud2019,Chainakun2021a,Ashton2022}. An increase in the index $\gamma$ of the disk emissivity, $\epsilon(r)=r^{-\gamma}$, and the disk parameter, $\alpha (H/R)^2$, where $\alpha$ is the viscosity
parameter and $H/R$ is the disk scale-height ratio, can produce more high-frequency variability power at higher energies. Therefore, varying $\gamma$ and $\alpha (H/R)^2$ is analogous to changing the bending power-law baseline model (e.g., bending index $s$ and $f_{\rm b}$).

Furthermore, our results show that the PSD shape flattens out as the energy increases ($s_{\rm s} > s_{\rm h}$), indicating that harder X-rays vary more at higher frequencies. This is well consistent with what is commonly observed in AGN \citep{Nandra2001,Vaughan2003,McHardy2004,McHardy2005,Papadakis2004,Gonzalez2012, Ashton2022}. We find that the PSD normalization in both energy bands ($A_{\rm s}$ and $A_{\rm h}$) are strongly correlated with $F_{\rm var}$ (see Fig.~\ref{fig-norm}). This is expected since the fractional excess variance is the integrated area under the PSD curve. We also find that the normalization is moderately correlated with the bending index $s$. The PSD shape then flattens not only with an increase in energies, but also with a decrease in $A$. Since $s$ is also related to the accretion parameters such as $\alpha (H/R)^2$, this supports the framework that the PSD amplitude governed by $A$ is dependent on the properties of the accretion rate \citep{Georgakakis2021}. 

Piecing these results together (e.g., Fig.~\ref{fig-params-dis}--\ref{fig-sum}), we can infer that as the corona height increases, the observed luminosity increases, and the disk itself also evolves in such a way that produces more high-frequency variability power (PSD spectrum is more flat) but produces less overall variability power (smaller $F_{\rm var}$). It is not straightforward how an increase in corona height, with increasing observed luminosity, induces more variability power at high frequencies. Perhaps, the corona and the disk variability evolve separately. Note that the correlation does not always imply causation. Therefore, it might be more intuitive to think that, when the corona height increases, the accretion disk itself becomes relatively more active at the inner regions so producing more power at high frequencies, rather than what it would be directly caused by the corona evolution.

\cite{Emmanoulopoulos2016} fitted the reverberation echoes in the PSD data of AGN. By extracting IRAS~13224--3809 PSD in the 0.5--1 keV band using the combined {\it XMM-Newton} light curve, they found the bending frequency of $\sim 0.7 \times 10^{-4}$~Hz and the bending index of $\sim 2.26$. Although the soft bands used here are slightly different, our average $f_{\rm b, s}$ is comparable to \cite{Emmanoulopoulos2016}, whilst the average $s_{\rm s}$ is smaller. In fact, our $f_{\rm b, s}$ falls within the bending frequencies reported by \cite{Alston2019}, who used two bending power-law components to fit the PSD data of IRAS~13224--3809. The results, as well, agree with \cite{Alston2019} that a higher bending frequency is required for the harder energy band.

In addition, the constrained $f_{\rm b}$ in both energy bands has a relatively weak correlation with the corona height. This suggests that the characteristic bend times-scales are less dependent on the coronal geometry. By assuming the IRAS~13224--3809 mass to be of $2 \times 10^{6} M_{\odot}$, the $f_{\rm b}$ obtained here is comparable (i.e., within the same order of magnitude) to the observation-based bending frequency that scales with the black hole mass \citep{Papadakis2004,Gonzalez2012}.

Last but not least, the luminosity bin data show a hint of the Lorentzian peak shifting towards lower frequencies for higher luminosity, consistent with \cite{Alston2019}. The FWHM of the line, when present, is $\sim 10^{-4}$~Hz. The Lorentzian features may be present in some individual observations, but they cannot be easily distinguished from the oscillatory structure that can be described by reverberation. \cite{Chainakun2021b} found that the reverberation features in the AGN PSD profiles can potentially be retrieved using the machine learning (ML) technique, allowing the source height to be inferred accurately. Applying ML techniques to track changes in geometry of all current reverberating AGN in a systematic way using the PSD data is a subject of future work (Mankatwit et al., in preparatory).

\section{Conclusion}

By utilizing the PSD analysis of IRAS~13224--3809, we report the consistent trend of increasing lamp-post source height with increasing luminosity, and compared it with previous literature. We find that the PSD model that includes the effects of X-ray reverberation can reasonably explain the oscillatory structures seen in the PSD profiles. However, while our source height ($\sim 3$--$25 \ r_{\rm g}$) is comparable to those observed in the time-lag data \citep{Alston2020}, it is substantially larger when compared with the source locations indirectly implied using the energy-integrated spectra \citep{Jiang2022}. This provides further evidence that the lamp-post parameters inferred from reverberation data are not consistent with those inferred from the time-averaged spectral analysis.

The model shows that when the corona height increases, the source luminosity increases as well, while the source spectrum tends to be softer. In addition, the observed fractional excess variance reduces. For the model to explain the data, a drop in $F_{\rm var}$ requires a smaller PSD normalization ($A$) and a smaller PSD bending index ($s$). Furthermore, both $A$ and $s$ are certainly linked to the accretion phenomena. This means that as the lamp-post source moves away from the black hole, the accretion disk evolves, probably independently, in a manner that produces less X-ray variability power making PSD spectrum to flatten out towards high frequencies.

Also, the trend of the PSD slopes that flatten out as energy increases is clearly observed. The Lorentzian features in individual observations cannot be clearly distinguished after accounting for reverberation effects. We can only see a hint of shifting the Lorentzian peak to low frequencies for high luminosity using the luminosity-bin data. Interestingly, the FWHM of the line remains almost constant regardless of the luminosity bin. High quality data (i.e. high signal-to-noise ratio) and long observations will be required to place robust constraints on the evolution of the PSD parameters.

$\newline$

This work (Grant No. RGNS~64--118) was supported by Office of the Permanent Secretary, Ministry of Higher Education, Science, Research and Innovation (OPS MHESI), Thailand Science Research and Innovation (TSRI), and Suranaree University of Technology. PC thanks funding support from the NSRF via the Program Management Unit for Human Resources \& Institutional Development, Research and Innovation (grant number B16F640076). WL was supported in part by Srinakharinwirot University (grants no. 035/2565) and the National Astronomical Research Institute of Thailand (NARIT). JJ acknowledges support from the Leverhulme Trust, the Isaac Newton Trust and St Edmund's College, University of Cambridge.

\bibliography{references}{}

\begin{thebibliography}{}
\expandafter\ifx\csname natexlab\endcsname\relax\def\natexlab#1{#1}\fi
\providecommand{\url}[1]{\href{#1}{#1}}
\providecommand{\dodoi}[1]{doi:~\href{http://doi.org/#1}{\nolinkurl{#1}}}
\providecommand{\doeprint}[1]{\href{http://ascl.net/#1}{\nolinkurl{http://ascl.net/#1}}}
\providecommand{\doarXiv}[1]{\href{https://arxiv.org/abs/#1}{\nolinkurl{https://arxiv.org/abs/#1}}}

\bibitem[{{Alston} {et~al.}(2019){Alston}, {Fabian}, {Buisson}, {Kara},
  {Parker}, {Lohfink}, {Uttley}, {Wilkins}, {Pinto}, {De Marco}, {Cackett},
  {Middleton}, {Walton}, {Reynolds}, {Jiang}, {Gallo}, {Zogbhi}, {Miniutti},
  {Dovciak}, \& {Young}}]{Alston2019}
{Alston}, W.~N., {Fabian}, A.~C., {Buisson}, D.~J.~K., {et~al.} 2019, \mnras,
  482, 2088, \dodoi{10.1093/mnras/sty2527}

\bibitem[{{Alston} {et~al.}(2020){Alston}, {Fabian}, {Kara}, {Parker},
  {Dovciak}, {Pinto}, {Jiang}, {Middleton}, {Miniutti}, {Walton}, {Wilkins},
  {Buisson}, {Caballero-Garcia}, {Cackett}, {De Marco}, {Gallo}, {Lohfink},
  {Reynolds}, {Uttley}, {Young}, \& {Zogbhi}}]{Alston2020}
{Alston}, W.~N., {Fabian}, A.~C., {Kara}, E., {et~al.} 2020, Nature Astronomy,
  4, 597, \dodoi{10.1038/s41550-019-1002-x}

\bibitem[{{Ar{\'e}valo} \& {Uttley}(2006)}]{Arevalo2006}
{Ar{\'e}valo}, P., \& {Uttley}, P. 2006, \mnras, 367, 801,
  \dodoi{10.1111/j.1365-2966.2006.09989.x}

\bibitem[{{Ashton} \& {Middleton}(2022)}]{Ashton2022}
{Ashton}, D.~I., \& {Middleton}, M.~J. 2022, arXiv e-prints, arXiv:2204.10346.
\newblock \doarXiv{2204.10346}

\bibitem[{{Caballero-Garc{\'\i}a} {et~al.}(2018){Caballero-Garc{\'\i}a},
  {Papadakis}, {Dov{\v{c}}iak}, {Bursa}, {Epitropakis}, {Karas}, \&
  {Svoboda}}]{Caballero2018}
{Caballero-Garc{\'\i}a}, M.~D., {Papadakis}, I.~E., {Dov{\v{c}}iak}, M.,
  {et~al.} 2018, \mnras, 480, 2650, \dodoi{10.1093/mnras/sty1990}

\bibitem[{{Caballero-Garc{\'\i}a} {et~al.}(2020){Caballero-Garc{\'\i}a},
  {Papadakis}, {Dov{\v{c}}iak}, {Bursa}, {Svoboda}, \& {Karas}}]{Caballero2020}
---. 2020, \mnras, 498, 3184, \dodoi{10.1093/mnras/staa2554}

\bibitem[{{Cackett} {et~al.}(2014){Cackett}, {Zoghbi}, {Reynolds}, {Fabian},
  {Kara}, {Uttley}, \& {Wilkins}}]{Cackett2014}
{Cackett}, E.~M., {Zoghbi}, A., {Reynolds}, C., {et~al.} 2014, \mnras, 438,
  2980, \dodoi{10.1093/mnras/stt2424}

\bibitem[{{Chainakun}(2019)}]{Chainakun2019a}
{Chainakun}, P. 2019, \apj, 878, 20, \dodoi{10.3847/1538-4357/ab1f0a}

\bibitem[{{Chainakun} {et~al.}(2022){Chainakun}, {Fongkaew}, {Hancock}, \&
  {Young}}]{Chainakun2022}
{Chainakun}, P., {Fongkaew}, I., {Hancock}, S., \& {Young}, A.~J. 2022, \mnras,
  513, 648, \dodoi{10.1093/mnras/stac924}

\bibitem[{{Chainakun} {et~al.}(2021{\natexlab{a}}){Chainakun}, {Luangtip},
  {Young}, {Thongkonsing}, \& {Srichok}}]{Chainakun2021a}
{Chainakun}, P., {Luangtip}, W., {Young}, A.~J., {Thongkonsing}, P., \&
  {Srichok}, M. 2021{\natexlab{a}}, \aap, 645, A99,
  \dodoi{10.1051/0004-6361/202039090}

\bibitem[{{Chainakun} {et~al.}(2021{\natexlab{b}}){Chainakun}, {Mankatwit},
  {Thongkonsing}, \& {Young}}]{Chainakun2021b}
{Chainakun}, P., {Mankatwit}, N., {Thongkonsing}, P., \& {Young}, A.~J.
  2021{\natexlab{b}}, \mnras, 506, 5318, \dodoi{10.1093/mnras/stab2098}

\bibitem[{{Chainakun} {et~al.}(2016){Chainakun}, {Young}, \&
  {Kara}}]{Chainakun2016}
{Chainakun}, P., {Young}, A.~J., \& {Kara}, E. 2016, \mnras, 460, 3076,
  \dodoi{10.1093/mnras/stw1105}

\bibitem[{{Chiang} {et~al.}(2015){Chiang}, {Walton}, {Fabian}, {Wilkins}, \&
  {Gallo}}]{Chiang2015}
{Chiang}, C.-Y., {Walton}, D.~J., {Fabian}, A.~C., {Wilkins}, D.~R., \&
  {Gallo}, L.~C. 2015, \mnras, 446, 759, \dodoi{10.1093/mnras/stu2087}

\bibitem[{{Churazov} {et~al.}(2001){Churazov}, {Gilfanov}, \&
  {Revnivtsev}}]{Churazov2001}
{Churazov}, E., {Gilfanov}, M., \& {Revnivtsev}, M. 2001, \mnras, 321, 759,
  \dodoi{10.1046/j.1365-8711.2001.04056.x}

\bibitem[{Dem\v{s}ar {et~al.}(2013)Dem\v{s}ar, Curk, Erjavec, Gorup,
  Ho\v{c}evar, Milutinovi\v{c}, Mo\v{z}ina, Polajnar, Toplak, Stari\v{c},
  \v{S}tajdohar, Umek, \v{Z}agar, \v{Z}bontar, \v{Z}itnik, \&
  Zupan}]{Demsar2013}
Dem\v{s}ar, J., Curk, T., Erjavec, A., {et~al.} 2013, J. Mach. Learn. Res., 14,
  2349–2353

\bibitem[{{Emmanoulopoulos} {et~al.}(2014){Emmanoulopoulos}, {Papadakis},
  {Dov{\v{c}}iak}, \& {McHardy}}]{Emmanoulopoulos2014}
{Emmanoulopoulos}, D., {Papadakis}, I.~E., {Dov{\v{c}}iak}, M., \& {McHardy},
  I.~M. 2014, \mnras, 439, 3931, \dodoi{10.1093/mnras/stu249}

\bibitem[{{Emmanoulopoulos} {et~al.}(2016){Emmanoulopoulos}, {Papadakis},
  {Epitropakis}, {Pech{\'a}{\v{c}}ek}, {Dov{\v{c}}iak}, \&
  {McHardy}}]{Emmanoulopoulos2016}
{Emmanoulopoulos}, D., {Papadakis}, I.~E., {Epitropakis}, A., {et~al.} 2016,
  \mnras, 461, 1642, \dodoi{10.1093/mnras/stw1359}

\bibitem[{{Epitropakis} {et~al.}(2016){Epitropakis}, {Papadakis},
  {Dov{\v{c}}iak}, {Pech{\'a}{\v{c}}ek}, {Emmanoulopoulos}, {Karas}, \&
  {McHardy}}]{Epitropakis2016}
{Epitropakis}, A., {Papadakis}, I.~E., {Dov{\v{c}}iak}, M., {et~al.} 2016,
  \aap, 594, A71, \dodoi{10.1051/0004-6361/201527748}

\bibitem[{{Fabian} {et~al.}(2013){Fabian}, {Kara}, {Walton}, {Wilkins}, {Ross},
  {Lozanov}, {Uttley}, {Gallo}, {Zoghbi}, {Miniutti}, {Boller}, {Brandt},
  {Cackett}, {Chiang}, {Dwelly}, {Malzac}, {Miller}, {Nardini}, {Ponti},
  {Reis}, {Reynolds}, {Steiner}, {Tanaka}, \& {Young}}]{Fabian2013}
{Fabian}, A.~C., {Kara}, E., {Walton}, D.~J., {et~al.} 2013, \mnras, 429, 2917,
  \dodoi{10.1093/mnras/sts504}

\bibitem[{{Fanton} {et~al.}(1997){Fanton}, {Calvani}, {de Felice}, \&
  {Cadez}}]{Fanton1997}
{Fanton}, C., {Calvani}, M., {de Felice}, F., \& {Cadez}, A. 1997, \pasj, 49,
  159, \dodoi{10.1093/pasj/49.2.159}

\bibitem[{Feigelson \& Babu(2012)}]{Feigelson2012}
Feigelson, E.~D., \& Babu, G.~J. 2012, Modern Statistical Methods for
  Astronomy: With R Applications (Cambridge University Press),
  \dodoi{10.1017/CBO9781139015653}

\bibitem[{{Georgakakis} {et~al.}(2021){Georgakakis}, {Papadakis}, \&
  {Paolillo}}]{Georgakakis2021}
{Georgakakis}, A., {Papadakis}, I., \& {Paolillo}, M. 2021, \mnras, 508, 3463,
  \dodoi{10.1093/mnras/stab2818}

\bibitem[{{George} \& {Fabian}(1991)}]{George1991}
{George}, I.~M., \& {Fabian}, A.~C. 1991, \mnras, 249, 352,
  \dodoi{10.1093/mnras/249.2.352}

\bibitem[{{Gonz{\'a}lez-Mart{\'\i}n} \& {Vaughan}(2012)}]{Gonzalez2012}
{Gonz{\'a}lez-Mart{\'\i}n}, O., \& {Vaughan}, S. 2012, \aap, 544, A80,
  \dodoi{10.1051/0004-6361/201219008}

\bibitem[{{Houck} \& {Denicola}(2000)}]{Houck2000}
{Houck}, J.~C., \& {Denicola}, L.~A. 2000, in Astronomical Society of the
  Pacific Conference Series, Vol. 216, Astronomical Data Analysis Software and
  Systems IX, ed. N.~{Manset}, C.~{Veillet}, \& D.~{Crabtree}, 591

\bibitem[{{Ingram} \& {Done}(2011)}]{Ingram2011}
{Ingram}, A., \& {Done}, C. 2011, \mnras, 415, 2323,
  \dodoi{10.1111/j.1365-2966.2011.18860.x}

\bibitem[{{Jansen} {et~al.}(2001){Jansen}, {Lumb}, {Altieri}, {Clavel}, {Ehle},
  {Erd}, {Gabriel}, {Guainazzi}, {Gondoin}, {Much}, {Munoz}, {Santos},
  {Schartel}, {Texier}, \& {Vacanti}}]{Jansen2001}
{Jansen}, F., {Lumb}, D., {Altieri}, B., {et~al.} 2001, \aap, 365, L1,
  \dodoi{10.1051/0004-6361:20000036}

\bibitem[{{Jiang} {et~al.}(2022){Jiang}, {Dauser}, {Fabian}, {Alston}, {Gallo},
  {Parker}, \& {Reynolds}}]{Jiang2022}
{Jiang}, J., {Dauser}, T., {Fabian}, A.~C., {et~al.} 2022, arXiv e-prints,
  arXiv:2204.09908.
\newblock \doarXiv{2204.09908}

\bibitem[{{Jiang} {et~al.}(2018){Jiang}, {Parker}, {Fabian}, {Alston},
  {Buisson}, {Cackett}, {Chiang}, {Dauser}, {Gallo}, {Garc{\'\i}a}, {Harrison},
  {Lohfink}, {De Marco}, {Kara}, {Miller}, {Miniutti}, {Pinto}, {Walton}, \&
  {Wilkins}}]{Jiang2018}
{Jiang}, J., {Parker}, M.~L., {Fabian}, A.~C., {et~al.} 2018, \mnras, 477,
  3711, \dodoi{10.1093/mnras/sty836}

\bibitem[{{Kara} {et~al.}(2013){Kara}, {Fabian}, {Cackett}, {Miniutti}, \&
  {Uttley}}]{Kara2013}
{Kara}, E., {Fabian}, A.~C., {Cackett}, E.~M., {Miniutti}, G., \& {Uttley}, P.
  2013, \mnras, 430, 1408, \dodoi{10.1093/mnras/stt024}

\bibitem[{{Karas} {et~al.}(1992){Karas}, {Vokrouhlicky}, \&
  {Polnarev}}]{Karas1992}
{Karas}, V., {Vokrouhlicky}, D., \& {Polnarev}, A.~G. 1992, \mnras, 259, 569,
  \dodoi{10.1093/mnras/259.3.569}

\bibitem[{{Lyubarskii}(1997)}]{Lyubarskii1997}
{Lyubarskii}, Y.~E. 1997, \mnras, 292, 679, \dodoi{10.1093/mnras/292.3.679}

\bibitem[{{Mahmoud} {et~al.}(2019){Mahmoud}, {Done}, \& {De
  Marco}}]{Mahmoud2019}
{Mahmoud}, R.~D., {Done}, C., \& {De Marco}, B. 2019, \mnras, 486, 2137,
  \dodoi{10.1093/mnras/stz933}

\bibitem[{{McHardy} {et~al.}(2005){McHardy}, {Gunn}, {Uttley}, \&
  {Goad}}]{McHardy2005}
{McHardy}, I.~M., {Gunn}, K.~F., {Uttley}, P., \& {Goad}, M.~R. 2005, \mnras,
  359, 1469, \dodoi{10.1111/j.1365-2966.2005.08992.x}

\bibitem[{{McHardy} {et~al.}(2004){McHardy}, {Papadakis}, {Uttley}, {Page}, \&
  {Mason}}]{McHardy2004}
{McHardy}, I.~M., {Papadakis}, I.~E., {Uttley}, P., {Page}, M.~J., \& {Mason},
  K.~O. 2004, \mnras, 348, 783, \dodoi{10.1111/j.1365-2966.2004.07376.x}

\bibitem[{{Miniutti} \& {Fabian}(2004)}]{Miniutti2004}
{Miniutti}, G., \& {Fabian}, A.~C. 2004, \mnras, 349, 1435,
  \dodoi{10.1111/j.1365-2966.2004.07611.x}

\bibitem[{{Nandra} \& {Papadakis}(2001)}]{Nandra2001}
{Nandra}, K., \& {Papadakis}, I.~E. 2001, \apj, 554, 710,
  \dodoi{10.1086/321423}

\bibitem[{{Papadakis} {et~al.}(2016){Papadakis}, {Pech{\'a}{\v{c}}ek},
  {Dov{\v{c}}iak}, {Epitropakis}, {Emmanoulopoulos}, \&
  {Karas}}]{Papadakis2016}
{Papadakis}, I., {Pech{\'a}{\v{c}}ek}, T., {Dov{\v{c}}iak}, M., {et~al.} 2016,
  \aap, 588, A13, \dodoi{10.1051/0004-6361/201527246}

\bibitem[{{Papadakis}(2004)}]{Papadakis2004}
{Papadakis}, I.~E. 2004, \mnras, 348, 207,
  \dodoi{10.1111/j.1365-2966.2004.07351.x}

\bibitem[{{Parker} {et~al.}(2017){Parker}, {Alston}, {Buisson}, {Fabian},
  {Jiang}, {Kara}, {Lohfink}, {Pinto}, \& {Reynolds}}]{Parker2017}
{Parker}, M.~L., {Alston}, W.~N., {Buisson}, D.~J.~K., {et~al.} 2017, \mnras,
  469, 1553, \dodoi{10.1093/mnras/stx945}

\bibitem[{{Parker} {et~al.}(2021){Parker}, {Alston}, {H{\"a}rer}, {Igo},
  {Joyce}, {Buisson}, {Chainakun}, {Fabian}, {Jiang}, {Kosec}, {Matzeu},
  {Pinto}, {Xu}, \& {Zaidouni}}]{Parker2021}
{Parker}, M.~L., {Alston}, W.~N., {H{\"a}rer}, L., {et~al.} 2021, \mnras, 508,
  1798, \dodoi{10.1093/mnras/stab2434}

\bibitem[{{Ponti} {et~al.}(2010){Ponti}, {Gallo}, {Fabian}, {Miniutti},
  {Zoghbi}, {Uttley}, {Ross}, {Vasudevan}, {Tanaka}, \& {Brandt}}]{Ponti2010}
{Ponti}, G., {Gallo}, L.~C., {Fabian}, A.~C., {et~al.} 2010, \mnras, 406, 2591,
  \dodoi{10.1111/j.1365-2966.2010.16852.x}

\bibitem[{{Remillard} \& {McClintock}(2006)}]{Remillard2006}
{Remillard}, R.~A., \& {McClintock}, J.~E. 2006, \araa, 44, 49,
  \dodoi{10.1146/annurev.astro.44.051905.092532}

\bibitem[{{Ross} \& {Fabian}(2005)}]{Ross2005}
{Ross}, R.~R., \& {Fabian}, A.~C. 2005, \mnras, 358, 211,
  \dodoi{10.1111/j.1365-2966.2005.08797.x}

\bibitem[{{Ross} {et~al.}(1999){Ross}, {Fabian}, \& {Young}}]{Ross1999}
{Ross}, R.~R., {Fabian}, A.~C., \& {Young}, A.~J. 1999, \mnras, 306, 461,
  \dodoi{10.1046/j.1365-8711.1999.02528.x}

\bibitem[{{Uttley} {et~al.}(2014){Uttley}, {Cackett}, {Fabian}, {Kara}, \&
  {Wilkins}}]{Uttley2014}
{Uttley}, P., {Cackett}, E.~M., {Fabian}, A.~C., {Kara}, E., \& {Wilkins},
  D.~R. 2014, \aapr, 22, 72, \dodoi{10.1007/s00159-014-0072-0}

\bibitem[{{Vaughan} {et~al.}(2003){Vaughan}, {Fabian}, \&
  {Nandra}}]{Vaughan2003}
{Vaughan}, S., {Fabian}, A.~C., \& {Nandra}, K. 2003, \mnras, 339, 1237,
  \dodoi{10.1046/j.1365-8711.2003.06285.x}

\bibitem[{{Wilkins} \& {Fabian}(2013)}]{Wilkins2013}
{Wilkins}, D.~R., \& {Fabian}, A.~C. 2013, \mnras, 430, 247,
  \dodoi{10.1093/mnras/sts591}

\bibitem[{{Wilkins} {et~al.}(2020){Wilkins}, {Garc{\'\i}a}, {Dauser}, \&
  {Fabian}}]{Wilkins2020}
{Wilkins}, D.~R., {Garc{\'\i}a}, J.~A., {Dauser}, T., \& {Fabian}, A.~C. 2020,
  \mnras, 498, 3302, \dodoi{10.1093/mnras/staa2566}

\bibitem[{{Zoghbi} {et~al.}(2020){Zoghbi}, {Kalli}, {Miller}, \&
  {Mizumoto}}]{Zoghbi2020}
{Zoghbi}, A., {Kalli}, S., {Miller}, J.~M., \& {Mizumoto}, M. 2020, \apj, 893,
  97, \dodoi{10.3847/1538-4357/ab7dc8}

\bibitem[{{Zoghbi} {et~al.}(2021){Zoghbi}, {Miller}, \& {Cackett}}]{Zoghbi2021}
{Zoghbi}, A., {Miller}, J.~M., \& {Cackett}, E. 2021, \apj, 912, 42,
  \dodoi{10.3847/1538-4357/abebd9}

\end{thebibliography}
\bibliographystyle{aasjournal}

\end{document}